\documentclass[showpacs,preprintnumbers,showkeys,twocolumn,prb,floatfix]{revtex4}
\usepackage{epsfig}
\usepackage{amsmath}

\newcommand{\figWidth}{8cm}
\newcommand{\inThesis}[1]{}
\newcommand{\inPaper}[1]{#1}
\input rotate

\newcommand{\rhov}{\mbox{\boldmath$\rho$}}
\newcommand{\epsilonv}{\mbox{\boldmath$\epsilon$}}
\begin{document}

\title{Phonons and Conduction in Molecular Quantum Dots: Density Functional Calculation of
  Franck-Condon Emission Rates for bi-fullerenes in External Fields}
\date{\today}

\author{Connie Te-ching Chang, James P. Sethna, Abhay N. Pasupathy, J. Park,
  D.C. Ralph, P.L. McEuen}
\affiliation{Laboratory of Atomic and Solid State Physics (LASSP), Clark Hall,
Cornell University, Ithaca, NY 14853-2501, USA}
\date{\today}

\begin{abstract}
We report the calculation of various phonon overlaps and their
corresponding phonon emission probabilities for the problem of an electron
tunneling onto and off of
the buckyball-dimer molecular quantum dot $C_{72}$, both with and without the influence of an external field.  We
show that the stretch mode of the two balls of the dumbbell couples most
strongly to the electronic transition, and in turn that a field in the direction of the bond between the two $C_{36}$ balls
is most effective at further increasing the phonon emission into the stretch mode.  As the field is increased,
phonon emission increases in probability with an accompanying decrease in
probability of the dot remaining in the ground vibrational state.  We also
present a simple model to gauge the effect of molecular size on the phonon emission of molecules similar to our $C_{72}$ molecule, including
the experimentally tested $C_{140}$.  In our approach we do not
assume that the hessians of the molecule are identical for different charge
states.  Our treatment is hence a generalization of the traditional 
phonon overlap calculations for coupled electron-photon transition in solids.     
\end{abstract}

 
\maketitle

\def \EInit {E_{1}}
\def \EFinal {E_{2}}
\def \KInit {K_{1}}
\def \KFinal {K_{2}}
\def \OmegaInit {\Omega_{1}}
\def \OmegaFinal {\Omega_{2}}
\def \OmegaBar {{{\Omega_{1} + \Omega_{2}} \over 2}}
\def \rFinal {{{\bf r}_{2}}}
\def \rInit {{{\bf r}_{1}}}
\def \F {{\bf F}}
\def \PsiInit {\Psi^{4-}}
\def \PsiFinal {\Psi^{5-}}
\def \y {{\bf y}}
\def \NInit {N_{1}}
\def \NFinal {N_{2}}
\def \NBar {{N_{{{1}+{2}}\over 2}}}
\def \tD {{\tilde\Delta}}
\def \qTA {{{\bf q}^{(5-)}_\alpha}}
\def \qBB {{{\bf q}^{({{2+3+}\over2})}_\beta}}

\section{Introduction}
Physics is full of examples of phonon-coupled quantum tunneling events.  A
classic example from the 1960s is the work done with trapped-electron color
centers in the lattices of the alkali-halides~\cite{schulman}.  More modern
examples include the study of how the mobility of interstitials in metals is modulated by
coupling of the defect to the resulting  
distortion of the surrounding lattice~\cite{flynn} and the study of how the
inter-chain hopping by polarons is affected by 
phonon interactions~\cite{conwell2}.  In these studies and others, the
frequencies before and after the transition were assumed to be unchanged and only
the coordinate about which the harmonic potential is centered shifts.  Here, our use
of the word {\it{phonon}}, traditionally used for plane-wave-like solutions in 
periodic
crystals, for {\it{vibrational normal mode}} is in the same spirit in this
context for
which we use the term {\it{quantum dot}}, a macroscale object, for {\it{molecule}}.
 
Over the past several years, several experiments and theoretical
studies~\cite{braig, mitra, schoeller} have been done where single
molecules have been used as the medium for vibration-assisted tunneling.
Some
recent experimental examples are measurements done with scanning tunnel microscopes
\cite{ho,stipe}, studies of single hydrogen molecules in mechanical break junctions
\cite{smit}, and the investigations that have directly motivated this work, the three-terminal
single molecule transistor experiments~\cite{park1,park2} where a single molecule is deposited
between two leads and is subjected to both a source-drain and gate bias.  This
is done in the Coulomb blockage regime, where the bias is tuned so that
sequential transport can occur and a differential conductance graph can be 
plotted.  In many
of these differential conductance graphs, in addition to the main lines due to
the change in the charge state of the molecule, there are a series of sidebands
thought to be caused by the coupling of the electron to the vibrational modes of
the molecule.  

\begin{figure}
\unitlength1cm
\begin{picture}(11.8,4.1)
\put(-7.2,-5.8){\makebox(15,11){
\includegraphics{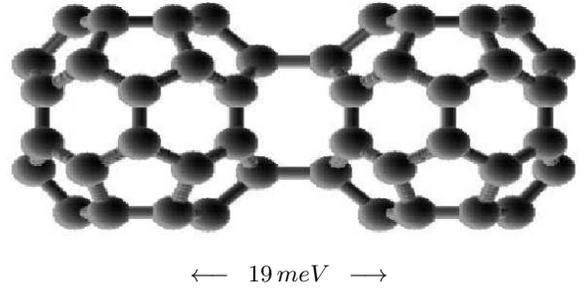}
}}
\put(3.4,-.11){$\longleftarrow\,\,\,19 \,meV\,\,\,\longrightarrow$}
\end{picture}
\caption{$C_{72}$ with 19 meV stretch mode indicated} 
\label{c72}
\end{figure}

In this paper, we present a general theory for these vibrational
overlaps where the vibrational modes of both the initial and final electronic
states of the molecule are considered.  Charge dependent hessians and 
anharmonic potentials in the context of single molecule transistors have been
considered previously~\cite{KV05,WN05} where the molecule is assumed to have one
dominant mode in each electronic state.  In the field of chemical spectroscopy,
this topic has been addressed~\cite{N01} through a general consideration of the Franck
Condon factors with Duschinsky rotation~\cite{D37} and its refinements~\cite{KRLC93,CDLW98}, which allows for different frequencies
and eigenvectors between different charged states.  Our paper considers a realistic model
of an N-atom molecule with 3N possible modes (for example, the bi-fullerene
$C_{72}$ with $216$ possible modes, see figure (\ref{c72})) and allows the calculation of experimental
scenarios by combining our formulation with results from existing quantum
chemistry packages.     

Spectroscopy has long been utilized as a tool in both chemistry and physics to
study the properties and structure of atoms and molecules.  Different types of
spectroscopy are used for different aims; optical spectroscopy for example studies the interaction of
electromagnetic radiation with the sample while this paper addresses
differential tunneling spectroscopy.  Franck-Condon factors serve as a
very good tool for analyzing the absorption and emission band intensities
corresponding to
vibrational levels in atoms and molecules~\cite{schatz}.  Over the years, many
such molecular vibrational spectra have been calculated and cataloged using Franck-Condon
factors~\cite{coon,herzberg}.  Single molecule transistors offer an
opportunity to apply the Franck-Condon principles to a new system.  Because we
are dealing with single molecules, we can calculate (using quantum chemistry
packages) the full vibrational profile of both the initial and final electronic
states of the molecule, and thus calculate the Franck-Condon intensities
generally.

Franck was the first to postulate that an electronic transition can be
accompanied by a vibrational excitation using classical arguments
 \cite{franck}. Condon later duplicated the argument using quantum mechanics and
the Born-Oppenheimer approximation~\cite{condon}.  According to Condon, the
intensity of a particular transition can be determined by calculating the
transition dipole moment.  The electronic component can be factored out, leaving
the square of the phonon overlap to modulate the total transition.

In section II, we set up our Hamiltonian.  IN section III, we outline our
version of the calculations of the phonon overlap calculations iincluding the
'Duschinsky rotation'.  In section IV, we outline our DFT numerical methods.
Section VI calculates the zero-field overlaps. Section VII addresses the
overlaps in a field. Section VIII introduces a simple two-ball and spring model
for the behavior of the stretch mode overlap in the presence of a field.
Section IX makes contact with $dI/dV$ measurements of the entire spectrum and
section X concludes. 
\section{The Hamiltonian}
\label{sec:HamiltoniansC72}
The Hamiltonian for the molecular dot looks like, in a mixed first and second quantized
formulation:
  \begin{equation}
  \label{e:Ham}
  H=H_1+H_2+H_3
  \end{equation}
where $H_1$, $H_2$, and $H_3$ are given by the following:
  \begin{equation}
  \label{first}
  H_1=\left[\sum_{k} \epsilon_k c^{l{\dagger}}_k c^{l}_k +  
  \sum_{k} \epsilon_k c^{r{\dagger}}_k c^{r}_k + 
  \epsilon_d c^{\dagger}_d c_d\right] 
  \end{equation}
  \begin{eqnarray}
  \label{second}
  H_2&=&{1\over{2}}{\bf{p}^{\dagger}\mathsf{M}^{-1}\bf{p}} \cr
  &+&(1-c^{\dagger}_d c_d) \left[1/2 (\bf{x}-\rInit)^\dagger 
  \mathsf{K_1}(\bf{x}-\rInit)\right]\cr 
  &+&c^{\dagger}_d c_d \left[1/2 (\bf{x}-\rFinal)^\dagger \mathsf{K_2}(\bf{x}-\rFinal)\right]  \end{eqnarray}
  \begin{equation} 
  \label{third}
  H_3=
   \sum_{k} T^{l}_k (c^{l{\dagger}}_k c_d +c^{\dagger}_d c^{l}_k) +\sum_{k} T^{r}_k
  (c^{r{\dagger}}_k c_d +c^{\dagger}_d c^{r}_k)
  \end{equation} 
Here we do not incorportate explicit terms for spin and charging effects because
we're focused on sequential tunneling events where only one elextron is on the
dot at a time.  Terms like the Coulomb effect in equation (\ref{beenakker})
below, we disregard in our expression.  
Equation~(\ref{first}) describes the electronic component involving 
the left and right leads and the dot, equation~(\ref{third}) describes the 
tunneling component, and equation~(\ref{second}) are the phonon states. 
Here $\bf{p}$ is the $3N$-component vector for the momentum of the N-atom molecule, $\mathsf{M}$ is the mass matrix (diagonal 
entries giving the masses of the different nuclei in groups of three),
$\bf{x}$ is the $3N$-dimensional vector for the atomic coordinates of the molecule, 
$\bf{r_{1}}$ and $\bf{r_{2}}$ are the $3N$-dimensional vectors for the minimum
energy configurations of the inital and final electronic states, 
and $\mathsf{K_{1}}$ and $\mathsf{K_{2}}$ are the quadratic forms giving the energy near
$\bf{r_1}$ and $\bf{r_2}$. The $1$ and $2$ indices reference the charge state of
the molecule.  In our transition, the $1$ index refers to the molecular
state with smaller charge, and the $2$ index refers to the higher
charge state.
$T^{l/r}_k$ is the tunneling matrix where the superscripts $l$ and $r$ 
specifies the left or right lead.  Finally, $c^\dagger$ and $c$ are creation 
and destruction operators, respectively, for electrons.
An external force $\bf{f}$ on the atomic coordinates
shifts the ground state configuration: e.g., 
$\bf{r_{2}} =  \bf{r^{(0)}_{2}} + \mathsf{K_2}^{-1}\bf{f}$.

Although the phonon states can also be expressed in
second quantized form via the creation and annihilation operators for bosonic
particles $a^{\dagger}$ and $a$, we chose to express them in first quantized
form to facilitate the calculation of the 3N-dimensional overlap integrals 
between different vibrational states of our initial and final molecule.  
\section{Phonon Overlap Integrals}
\label{sec:Overlap}
In regarding our problem, we are considering the case of incoherent tunneling,
where one electron is tunneling on or off the molecular quantum dot at a time.  We also assume that the molecule has time to relax to its minimum
energy configuration in between tunneling events such that all phonon
excitations decay before the next tunneling event.  In recent experiments
\cite{leroy}, long phonon lifetimes extending at least fifty times beyond the
lifetimes observed in Raman spectroscopy have been measured for experiments on
suspended carbon nanotubes.  However, the authors note that the lack of coupling
to a substrate may account for this increase.  In other experiments \cite{ho},
the experimental setup was arranged to increase the lifetime of the electron
compared to the phonon, allowing for observation of transient vibronic levels.
For our calculation, we presume that low currents and strong phonon coupling between molecule
and leads ensures vibrational relaxation between electron tunneling events.

Secondly, we assume the Born-Oppenheimer approximation where the total wave
function is described by
$|\Psi(z,x)\rangle=|\varphi_{x}(z)\phi(x)\rangle$, where $x$ labels the
nuclear coordinates as above, and $z$ labels the electron coordinates.
Strictly speaking, the ground-state electron wave function depends
parametrically on the nuclear positions $x$.  
Based on geometric optimization calculations, we know the atomic fluctuations $\delta x$ are on the order of
picometers, so we can safely assume $\phi_{x}(z)\approx \phi(z)$ and hence 
factor our wavefunction into a
purely electronic component and a purely nuclear component.  
The transition then is given by the following matrix elements where $T$ is
given by (\ref{third}):
  \begin{eqnarray}
  \label{MatrixElement}
  T_{fi}&=&\langle\Psi_f(z,x)|T|\Psi_i(z,x)\rangle\cr
 &=&\langle\varphi^{i}(z)\phi^{i}(x)|T|\varphi^{f}(z)\phi^{f}(x)\rangle \cr
  &=& \langle\varphi^{f}(z)|T|\varphi^{i}(z)\rangle
  \langle\phi^{f}(x)|\phi^{i}(x)\rangle  
  \end{eqnarray}

Following the Landauer approach~\cite{landauer,buttiker,Beenakker2} and using Fermi's Golden Rule to give us a
transition probability, we can find the conductance. 
Sequential tunneling in the Coulomb Blockade regime is assumed 
~\cite{Beenakker} for the case where the quantum dot has energy
levels given by $E_{p}$ (For us, $E_p$ will represent one of many energy
eigenstates with both a change in electron charge and the excitation of one or
more vibrational modes). The 
formula for the conductance is then found to be:
  \begin{eqnarray}
  \label{beenakker}
  conductance&=&{e^{2}\over{kT}}\sum_{p}\sum_{N}
    \Big\{{\Gamma^{l}_{p}\Gamma^{r}_{p}\over{\Gamma^{l}_{p}+\Gamma^{r}_{p}}}
    P_{eq}(N)F_{eq}(E_{p}|N)\cr
& &\times[1-f(E_{p}+U(N)-U(N-1)-E_{F})]\Big\}
  \end{eqnarray}
where $\Gamma^{l/r}_{p}$ is the tunneling rate from
the dot electron energy level $p$ to the left/right leads, 
$N$ is the number of electrons in the dot before the tunneling event, 
$U(N)$ is the charging energy for $N$ electrons on the dot,
$F_{eq}(E_{p}|N)$ is the conditional
probability in equilibrium that the level p is occupied given that the quantum
dot contains $N$ electrons, $E_F$ is the chemical potential of the leads, and $P_{eq}(N)$ is the probability that the quantum
dot has $N$ electrons in equilibrium.  In addition, the electrons in the
leads are in the Fermi distribution $f(E)$. 

Since tunneling rates depend
exponentially on separation, the tunneling rate through one of the leads, 
say the right one, is often much smaller
than the other, with the smaller rate acting as the bottleneck, 
${\Gamma^{l}_{p}\Gamma^{r}_{p}\over{\Gamma^{l}_{p}+\Gamma^{r}_{p}}} 
 \sim \Gamma^r_p$.
Hence the dependence of $\Gamma^r_p$ on the final state $p$ determines the
variation of the conductance with energy. This tunneling rate is proportional
to $|T_{fi}|^2$, 
where $|T_{fi}|$ is given by (\ref{MatrixElement}). For a given electronic
transition $\varphi^i$ to $\varphi^f$, assuming (at the low temperatures
of these experiments) that the initial molecular state has no phonon
excitations, the dependence of this matrix element on the final phonon state
is given primarily by the phonon overlap integrals in equation~(\ref{MatrixElement}),
and hence
  \begin{equation}
  \label{overlapintegral}
  G \propto \Gamma^r_p \propto \sum_{\epsilon_F\leq eV}|\langle\phi^{f}(x)|\phi^{i}(x)\rangle|^2.
  \end{equation}

The linear conductance given in equation (\ref{overlapintegral}) is dependent only on
the ground state since the excitation of phonons is related to the applied
bias difference across the molecule.  As each threshold step in bias is crossed, new possible pathways are
accessible and the squares of their overlaps must be added to the expression.

This phonon overlap integral is the quantity of interest since it modulates
the total
transition rate.  Its value is a measure of the probability of occurrence of a
particular transition between the initial vibrational state of the initial
charge state (assumed to always
be the ground state) and the final vibrational state of the final charge state.
This quantity will suppress the total transition rate matrix element, leading to
less intensity in the line. Summing over final states yields 1~\cite{GS88}:
  \begin{equation}
  \label{sumone}
  \sum_{f}|\langle\phi^{f}(x)|\phi^{i}(x)\rangle|^{2}=1
  \end{equation}
with the individual terms representing the probability decomposition of the
initial state in the eigenstates of the final potential.  Hence the weight of
the original transition is spread among the phonon excitations.

\subsection{Normal modes and phonon wavefunctions}
\label{subsec:3Noverlaps}
In 3N dimensions, the phonon Hamiltonian for the initial charge state is:
  \begin{equation}
  \label{firstham}
  H={1\over{2}}{\bf{p}}^{\dagger}\mathsf{M}^{-1}{\bf{p}}+{1\over{2}}({\bf{x}}-\bf{r_1})^{\dagger}\mathsf{K_{1}}(\bf{x}-\bf{r_1})
  \end{equation}
For molecules with atoms of unequal mass, transforming from position space to
normal modes becomes much simpler if we use the standard trick of rescaling the
coordinates by the square root of the mass and shift the origin to
$\bf{r_1}$, the equilibrium configuration of the initial charge state:   
  \begin{equation}
  \label{eqn:position}
  \bf{y}=\mathsf{M}^{1/2}(\bf{x}-\bf{r_1}).
  \end{equation}
Hence:
  \begin{equation}
   H_1 ={{\bf{\Pi}}^{\dagger}{\bf{\Pi}}\over{2}}+{1\over{2}}{\bf{y}}^{\dagger}\mathsf{\Omega_{1}}^{2}{\bf{y}}
  \end{equation}
where $\Pi=M^{-1}P$ and $\mathsf{\Omega}_{i}^{2}=\mathsf{M}^{-1/2}\mathsf{K}_{i}\mathsf{M}^{-1/2}$ is a matrix with dimensions of 
frequency squared.  

Similarly, the phonon Hamiltonian for the final charge state
is:
  \begin{eqnarray}
  H_{2}&=&{{\bf{p}}^{\dagger}\mathsf{M}^{-1}{\bf{p}}\over{2}}+{1\over{2}}({\bf{x}}-\rFinal)^{\dagger}\mathsf{K}_2({\bf{x}}-\rFinal)\cr
  &=&{1\over{2}}{\bf{\Pi}}^{\dagger}{\bf{\Pi}}+{1\over{2}}({\bf{y}}-{\bf{\Delta}})^{\dagger}\mathsf{\Omega}_{2}^{2}({\bf{y}}-{\bf{\Delta}})
  \end{eqnarray}
where:
  \begin{equation}
  \label{Deltaeq}
  {\bf{\Delta}}=\mathsf{M}^{1/2}(\rFinal-\rInit)
  \end{equation}
is the rescaled atomic displacement due to the change in charge state.
\inThesis{
We solve the Schrodinger equation in matrix form:
  \begin{equation}
  \label{eq:sch}
  H\Psi=E\Psi
  \end{equation}
The Hamiltonian in matrix form can be written as follows:
  \begin{equation}
  H=-{\hbar^{2}\over{2}}\left(
  \begin {array}{cccc}
  {\partial^{2}\over{\partial y_{1}^{2}}}&0&0&...\\
  \noalign{\medskip}
  0&{\partial^{2}\over{\partial y_{2}^{2}}}&0&...\\
  \noalign{\medskip}
  0&0&{\partial^{2}\over{\partial y_{3}^{2}}}&...\\
  \noalign{\medskip}
  ...&...&...&...
  \end {array}\right)
+{1\over{2}}\left(
  \begin {array}{cccc}
  \Omega_{11}^{2}y_{1}^{2}&\Omega_{12}^{2}y_{1}y_{2}&\Omega_{13}^{2}y_{1}y_{3}&...\\
  \noalign{\medskip}
  \Omega_{21}^{2}y_{2}y_{1}&\Omega_{22}^{2}y_{2}^{2}&\Omega_{23}^{2}y_{2}y_{3}&...\\
  \noalign{\medskip}
  \Omega_{31}^{2}y_{3}y_{1}&\Omega_{32}^{2}y_{3}y_{2}&\Omega_{33}^{2}y_{3}^{2}&...\\
  \noalign{\medskip}
  ...&...&...&...
  \end{array}
  \right)
  \end{equation}

We see that the kinetic energy part of the Hamiltonian is diagonal in the
coordinate basis, but that the harmonic potential energy part  
of the Hamiltonian
is not.  We would like to diagonalize the entire Hamiltonian to make the
multi-dimensional Schrodinger equation separable and thus find the wave
function $\Psi$.
To make the potential part of the Hamiltonian diagonal, we need to transform
from the coordinate basis to the normal mode basis.  This will yield a
Hamiltonian for a series of one-dimensional simple harmonic oscillators.  

We can write our position vectors $y$ in terms of the orthonormal basis which
diagonalizes $\Omega_{1}$:
  \begin{equation}
  \label{eq:normal}
  \vec{y}=\sum_{i}q_{i}\hat{\epsilon^{i}}
  \end{equation}
where $q_{i}$ are the coefficients weighting the eigenvectors 
$\{\hat{\epsilon}\}$ and the index $i$ indicates the mode
number. The potential part of the Hamiltonian then becomes:
  \begin{equation}
  {1\over{2}}\left(
  \begin {array}{cccc}
  \omega_{1}^{2}q_{1}^{2}&0&0&...\\
  \noalign{\medskip}
  0&\omega_{2}^{2}q_{2}^{2}&0&...\\
  \noalign{\medskip}
  0&0&\omega_{3}^{2}q_{3}^{2}&...\\
  \noalign{\medskip}
  ...&...&...&...
  \end{array}
  \right)
  \end{equation}

Now, to ensure that the entire Hamiltonian is diagonal, we need to make sure
that the kinetic energy part remains diagonal when transformed to the new
basis from the position coordinate basis set.  We start with
${\partial^{2}\Psi\over{\partial y_{i}^{2}}}$ and show that this is equal to
${\partial^{2}\Psi\over{\partial q_{i}^{2}}}$.

Starting from our definition of the transformation from position coordinate
basis to normal coordinate basis:
  \begin{eqnarray}
  \label{eq:coorb}
  \vec{y}&=&\sum_{i}q_{i}\hat{\epsilon}^{i}\cr
  \vec{y}&=&E\cdot{\vec{q}}
  \end{eqnarray}
ith $E_{ij}=(\hat{\epsilon}^{i})_{j}$, the $j^{th}$ component of the $i^{th}$
eigenfrequency, we can then write:
  \begin{eqnarray}
  \label{eq:proof}
  {\partial\Psi\over{\partial y_{i}}}&=&{\partial\Psi\over{\partial
      q_{j}}}\cdot{{\partial q_{j}\over{\partial y_{i}}}}\cr
  &=&{\partial\Psi\over{\partial q_{j}}}\cdot{E_{ji}}.
  \end{eqnarray}
Taking the second derivative, we get:
  \begin{eqnarray}
  \label{eq:secderiv}
  {\partial^{2}\Psi\over{\partial y_{i} \partial y_{l}}}&=&{\partial\over{\partial
      q_{k}}}({\partial\Psi\over{\partial q_{j}}} E_{ji})\cdot{{\partial
      q_{k}\over{\partial{y_{l}}}}}\cr
  &=&{\partial^{2}\Psi\over{\partial q_{k} \partial q_{j}}}E_{ji}E_{kl}
  \end{eqnarray}
Substituting $i=l$, we get and using the fact that the normal modes are orthonormal:
  \begin{eqnarray}
  \label{equation:fi}
  \bigtriangledown^{2}_{r}\Psi&=&{\partial^{2}\Psi\over{\partial
    q_{k} \partial q_{j}}}E_{ji}E_{ki}={\partial^{2}\Psi\over{\partial
    q_{k} \partial q_{j}}}\hat{\epsilon^{j}}\cdot{\hat{\epsilon^{k}}}\cr
&=&{\partial^{2}\Psi\over{\partial
    q_{k} \partial q_{j}}}\delta_{jk}
=\bigtriangledown_{q}^{2}\Psi.
  \end{eqnarray}
Note that the transformation to normal modes is only orthonormal in the 
rescaled coordinates (for atoms of unequal mass). 

Our Hamiltonian is now fully diagonal in the phonon mode basis, the
Schrodinger equation is now separable, and the ground state solution $\Psi$ is the product of
the ground state solution $\psi_{i}$s for the one-dimensional harmonic oscillator:
  \begin{eqnarray}
  \label{eq:sho}
  \Psi_{0}&=&\prod_{i}\psi_{i}\cr
  \Psi_{0}&=&\prod_{i}\left({\omega_{i}\over{\pi\hbar}}\right)^{1/4}e^{-{1\over{2\hbar}}(w_{i}q_{i}^{2})}\sim
  e^{\sum_{i}-{1\over{2\hbar}}(\omega_{i}q_{i}^{2})}.
  \end{eqnarray}    
Now, we transform our basis back into the position coordinate basis using the
definition $\vec{y}=\sum_{k} q_{k} \hat{\epsilon_{k}}.$
Reversing the transform, we get:
  \begin{eqnarray}
  \label{eq:transform}
  \vec{r\y}\cdot{\hat{\epsilon_{i}}}=\sum_{k}q_{k}\hat{\epsilon_{k}}\cdot{\hat{\epsilon_{i}}}\cr
  \rightarrow q_{i}=\vec{y}\cdot{\hat{\epsilon}_{i}}.
  \end{eqnarray}
The wavefunction then becomes:
  \begin{eqnarray}
  \label{eq:what}
  \Psi_{0}&\sim& e^{\sum_{i}-{1\over{2\hbar}}
    \omega_{i}(\vec{y}\cdot{\hat{\epsilon}_{i}})^{2}}\cr
  &\sim& e^{-{1\over{2\hbar}}\omega_{i} (\vec{y}\cdot{\hat{\epsilon}_{i}})(\vec{y}\cdot{\hat{\epsilon}_{i}})^{T}}\cr
  &\sim& e^{-{1\over{2\hbar}} y^{T} E^{T}\Lambda E y}\cr
  &\sim& e^{-{1\over{2\hbar}} y^{T} \Omega y}
  \end{eqnarray}
where $\Lambda$ is a diagonal matrix with the frequencies of 
the simple harmonic
oscillator $\omega_{i}$ for each mode $i$ along the diagonal and using 
the fact that
multiplying the normal mode eigenvectors on either side of this diagonal matrix
yields the matrix $\Omega$.  See also the following section
(\ref{normalmodebasis}) for an equivalent derivation.

Now, we can normalize our wavefunction by integrating over all degrees of
freedom.  In the normal mode basis
$A^{2}\Pi_{k}({\pi\hbar\over{\omega_{k}}})^{1/2}=1$.  In the real space basis, we have:
  \begin{eqnarray}
  \label{eq:norm}
  \int ... \int (Ae^{{1\over{2\hbar}} y^{T}\Omega y})(Ae^{{1\over{2\hbar}} y^{T}\Omega y}) d^{n}y &=& 1 \cr
  \int ... \int A^{2}e^{-{1\over{\hbar}}y^{T}\Omega y} d^{n}y &=& 1\cr
  A^{2}\left({\hbar^n \pi^{n}\over{\det(\Omega)}}\right)^{1/2} &=& 1\cr
  \rightarrow A=\left(\det\left({\Omega\over{\pi\hbar}}\right)\right)^{1/4}
  \end{eqnarray}
which is equal to $A = \prod_\alpha \left({\omega_\alpha \over \pi \hbar}\right){1/4}$ (as we found in the normal mode basis) because 
$\det \Omega = \det \Lambda = \prod_\alpha \omega_\alpha$.

Therefore our ground state wavefunction for our system is:
  \begin{equation}
  \label{ground}
  \Psi_{0}(y)= \left(\det\left({\Omega\over{\pi\hbar}}\right)\right)^{1/4}e^{-{1\over{2\hbar}} y^{T}
    \Omega y}
  \end{equation}
and excited wavefunctions are then given by
  \begin{equation}
  \label{excited}
  \Psi_{n,\alpha}(y) = 
	\left(\det\left({\Omega\over{\pi\hbar}}\right)\right)^{1/4}
   {H_{n}^{\alpha}\left(\sqrt{\omega_\alpha/\hbar}
			(y\cdot{{\hat \epsilon}_{\alpha}^{(n)}})\right)
	\over
   {2^{n/2}\sqrt{n!}}} 
	\, \exp\left(-{1\over{2\hbar}} y^T \Omega y \right)
  \end{equation}
where $y\cdot{{\hat \epsilon}_\alpha^{(n)}}$ represents the component of $y$
along the $n^{th}$ phonon eigenmode, and 
$y\cdot{{\hat \epsilon}_\alpha^{(1)}}=q_\alpha$.

\subsection{3N-dimensional wavefunctions and normalizations integrating in
		normal mode basis}
\label{normalmodebasis}
We can also find the above wavefunctions working in the normal mode basis. 
Again, starting from the expression of the 3-N dimensional wavefunction probability in the
coordinate basis, we need to integrate over all the degrees of freedom and set
the result equal to unity to find the proper normalization factor:
\begin{equation}
N^{2}\int{d\vec{y}e^{-{1\over{\hbar}}\vec{y}^{T}\Omega\vec{y}}}=1.
\end{equation}

Again, we define a change in basis from coordinate to normal modes:
\begin{equation}
\vec{y}=\sum_{\alpha}q_{\alpha}\hat{\epsilon}^{(\alpha)}
\end{equation}
where $\alpha$, the number of modes, runs over the dimensions of the problem (i.e., 3N for a three-dimensional
molecule with $N$ atoms).

Component by component, the above is expressed as:
\begin{equation}
\label{convert}
y_{i}=\sum_{\alpha}q_{\alpha}\hat{\epsilon}_{i}^{\alpha}.
\end{equation}
The $q_{\alpha}$'s are the coefficients in front of each eigenmode 
that describes the
linear combination of eigenmodes that make up the coordinate vector $\vec{y}$;
and $\hat{\epsilon}^{(\alpha)}$ is the $\alpha^{th}$ mode eigenvector.

In the normal mode basis, 
\begin{eqnarray}
\Omega=E \Lambda E^{T}\cr 
\Lambda=E^{T}\Omega E
\end{eqnarray}
where $E$ is the matrix of orthonormal eigenvectors of the matrix $\Omega$, or
in other words, $E$ is the matrix of the eigenvectors describing the modes of
the molecule.  As before $\Lambda$ is the diagonal matrix 
with diagonal elements
$\sim\omega^{2}$.

Now, we transform the quantity $\vec{y}^{T}\Omega\vec{y}$ into the normal mode
basis where we sum over the modes $\alpha$,$\beta$:
\begin{eqnarray}
\label{vector1}
\vec{y}^{T}\Omega\vec{y}&=&\sum_{\alpha,\beta}(q_{\alpha}\hat{\epsilon}^{(\alpha)^{T}})(E\Lambda
E^{T})(q_{\beta}\hat{\epsilon}^{(\beta)})\cr
&=&\sum_{\alpha,\beta}q_\alpha q_\beta(E^{T}\hat{\epsilon}^{(\alpha)})^{T}\Lambda(E^{T}\hat{\epsilon}^{(\beta)})\cr
&=&\sum_{\alpha,\beta}q_{\alpha}q_{\beta}(E^{T}\hat{\epsilon}^{\alpha})^{T}[\Lambda(E^{T}\hat{\epsilon}^{\beta})]
\end{eqnarray}
But the expression
$((E^{T}\hat{\epsilon}^{\alpha})^{T}[\Lambda(E^{T}\hat{\epsilon}^{\beta})]$ is
just a dot product which can be expressed in the form $\sum_{i}A_{i}B_{i}$.

After noting that
$E_{i\alpha}=\hat{\epsilon}^{\alpha}_{i}$ since the $i\alpha$ element of the
matrix $E$ is indexed by the column $\alpha$ which locates the $\alpha^{th}$
mode and the row $i$ which locates the $i^{th}$ component of the eigenvector, 
we can write the dot product as: 
\begin{eqnarray}
\label{dots}
\sum_{i}A_{i}B_{i}&=&\sum_{i}(E^{T}\hat{\epsilon}^{\alpha})_{i}[\Lambda(E^{T}\hat{\epsilon}^{\beta})_{i}]\cr
&=&\sum_{i}[E^{T}_{ij}\hat{\epsilon}_{j}^{\alpha}][\Lambda_{ik}(E^{T}\hat{\epsilon}^{\beta})_{k}]\cr
&=&\sum_{i}[E_{ji}\hat{\epsilon}_{j}^{\alpha}][\Lambda_{ik}(E_{k\ell}^{T}\hat{\epsilon}_{\ell}^{\beta})]\cr
&=&\sum_{i}[\hat{\epsilon}_{j}^{i}\hat{\epsilon}_{j}^{\alpha}][\Lambda_{ik}(E_{\ell k}\hat{\epsilon}_{\ell}^{\beta})]\cr
&=&\sum_{i}[\delta_{i\alpha}][\Lambda_{ik}(\hat{\epsilon}_{\ell}^{k}\hat{\epsilon}_{\ell}^{\beta})]\cr
&=&\sum_{i}\delta_{i\alpha}\Lambda_{ik}\delta_{k\beta}\cr
&=&\sum_{i}\delta_{i\alpha}\Lambda_{i\beta}\cr
&=&\Lambda_{\alpha\beta}.
\end{eqnarray}

Plugging in the result for (\ref{dots}) into the expression in
(\ref{vector1}) yields:
\begin{eqnarray}
\vec{y}^{T}\Omega\vec{y}&=&\sum_{\alpha\beta}q_{\alpha}q_{\beta}\sum_{i}(E^{T}\hat{\epsilon}^{\alpha})_{i}(\Lambda(E^{T}\hat{\epsilon}^{\beta}))_{i}\cr
&=&\sum_{\alpha\beta}q_{\alpha}q_{\beta}\Lambda_{\alpha\beta}\cr
&=&\sum_{\alpha}q_{\alpha}^{2}\omega_{\alpha}
\end{eqnarray}
where $\alpha$ runs over the modes.

Now, our integral in coordinate basis can be expressed in normal mode basis:
  \begin{equation}
  N^{2}\int{d\vec{y}}e^{-{1\over{\hbar}}\vec{y}^{T}\Omega\vec{y}}
    =N^{2}\int{d\vec{q}}
      e^{-{1\over{\hbar}}\sum_{\alpha}\omega_{\alpha}q_{\alpha}^{2}} J(\vec{q})
  \end{equation}
where $J(\vec{q})$ is the Jacobian of the transformation between basis.

When making a change in variable in an integral, the Jacobian of the
transformation is needed.  For example, for a change of variables from
$x,y\rightarrow u,v$ in the form $x=x(u,v)$ and $y=y(u,v)$, the Jacobian given
by:
  \begin{equation}
  J(u,v)=\left|
  \begin{array}{ c c}
  \partial{x}\over{\partial{u}}&\partial{x}\over{\partial{v}}\\
  \partial{y}\over{\partial{u}}&\partial{y}\over{\partial{v}}
  \end{array} \right|,
  \end{equation} 
and is needed to recast the integral in the new variables:
$\int{\int{dx}dy}f(x,y)=$$\newline$
$\int{\int{du}dv}|J(u,v)|f(x(u,v),y(u,v))$.

In our case, our Jacobian is:
  \begin{equation}
  J(\vec{q})=\left|
  \begin{array}{ c c c c }
  \partial{y_{1}}\over{\partial{q_{1}}}&\partial{y_{1}}\over{\partial{q_{2}}}&\partial{y_{1}}\over{\partial{q_{3}}}&...\\
  \partial{y_{2}}\over{\partial{q}_{1}}&\partial{y_{2}}\over{\partial{q_{2}}}&\partial_{y_{2}}\over{\partial{q_{3}}}&...\\
  \partial{y_{3}}\over{\partial{q_{1}}}&\partial{y_{3}}\over{\partial{q_{2}}}&\partial{y_{3}}\over{\partial{q_{3}}}&...\\
  ...&...&...&...\\
  \end{array} \right|.
  \end{equation} 
But referring to the definition given in (\ref{convert}), the Jacobian is just:
  \begin{equation}
  J(\vec{q})=\left|
  \begin{array}{ c c c c }
  {\hat \epsilon}_{1}^{(1)}&{\hat \epsilon}_{1}^{(2)}&{\hat \epsilon}_{1}^{(3)}...\\
  {\hat \epsilon}_{2}^{(1)}&{\hat \epsilon}_{2}^{(2)}&{\hat \epsilon}_{2}^{(3)}...\\
  {\hat \epsilon}_{3}^{(1)}&{\hat \epsilon}_{3}^{(2)}&{\hat \epsilon}_{3}^{(3)}...\\
  ...&...&...&...\\
  \end{array} \right|
  \end{equation} 
which is just the determinant of the matrix made up of the normal mode
eigenvectors $\hat\epsilon$. Since the normal modes are orthonormal,
this Jacobian equals one. 

Hence
the integral over all degrees of freedom for the ground state
vibrational wavefunction probability in normal mode coordinates is:
\begin{eqnarray}
N^{2}\int{d\vec{q}}e^{-{1\over{\hbar}}\sum_{\alpha}\omega_{\alpha}q_{\alpha}^2}=1\cr
N^{2}\int{dq_{1}}e^{-{1\over{\hbar}}\omega_{1}q_{1}^{2}}\int{dq_{2}}e^{-{1\over{\hbar}}\omega_{2}q_{2}^{2}}...\int{dq_{n}}e^{-{1\over{\hbar}}\omega_{n}q_{n}^{2}}=1\cr
N^{2}\sqrt{\pi\hbar\over{\omega_{1}}}\sqrt{\pi\hbar\over{\omega_{2}}}...\sqrt{\pi\hbar\over{\omega_{n}}}=1\cr
N=\sqrt[4]{\Pi_{i}{\omega_{i}\over{\pi\hbar}}}=\sqrt[4]{\det({\Lambda\over{\pi\hbar}})}\cr
N=\sqrt[4]{\det({\Omega\over{\pi\hbar}})}.
\end{eqnarray}
Our wavefunctions are then those given previously in equations~(\ref{ground}) and
(\ref{excited}).
}
\subsection{3N-dimensional wavefunctions and overlaps}
\label{subsec:3ndoverlaps}

In this section we calculate the transition rate from the neutral 
molecule's ground state to the ground state and the various singly
excited vibrational states of the charged molecule. Our calculation
of the Franck-Condon factors is thus the one-phonon emission special
case of the more general Duschinsky rotation calculations in the 
chemistry literature~\cite{KRLC93}. We present it here partly because
we find this special case physically illuminating, and partly to 
introduce our notation. We present in the Appendix the 
more complex calculation of the Franck-Condon factor from the neutral
ground state to a doubly-excited vibrational charged state.

Using the Hermite polynomials associated with solutions to the harmonic
oscillator ($H_1(x) = 2 x$ and $H_2(x) = -2 + 4 x^2$), and the expression
for the excited wavefunctions, we
have the 3N-dimensional vibrational eigenfunctions:
\begin{eqnarray}
\label{PsiGroundStates}
\Psi^{(1)}_{0}({\bf{y}}) &=& {{N}}_{1} e^{-{1\over{2\hbar}}{\bf{y}}^{\dagger} \mathsf{\Omega}_{1} {\bf{y}}}\cr
\Psi^{(2)}_0({\bf{y}}) &=& {{N}}_{2}e^{-{1\over{2\hbar}} ({\bf{y}}-{\bf{\Delta}})^{\dagger}\mathsf{\Omega}_2 ({\bf{y}}-{\bf{\Delta}})} \cr
\Psi^{(2)}_{1,\alpha}({\bf{y}}) &=& {N}_{2} \, \sqrt{2\omega_\alpha/\hbar} 
(({\bf{y}}-{\bf{\Delta}})\cdot {{\bf{\hat \epsilon}}}_{\alpha}^{(2)}) \cr
& &\times \exp\left(-{1\over{2\hbar}}({\bf{y}}-{\bf{\Delta}})^{\dagger}
\mathsf{\Omega}_{2} ({\bf{y}}-{\bf{\Delta}})\right)\cr
\Psi^{(2)}_{2,\alpha}({\bf{y}}) &=& {{{N}}_{2}\over{2 \sqrt{2}}} \, H_{2}(\sqrt{\omega_\alpha/\hbar} 
(({\bf{y}}-{\bf{\Delta}})\cdot {{\bf{\hat \epsilon}}}_{\alpha}^{(2)})) \cr
& &\times\exp\left(-{1\over{2\hbar}}({\bf{y}}-{\bf{\Delta}})^{\dagger} 
\mathsf{\Omega}_{2} ({\bf{y}}-{\bf{\Delta}})\right).
\end{eqnarray}

Here, the ${\bf{\Delta}}$ encapsulates the geometric reconfiguration of the molecule
(equation~\ref{Deltaeq}), the superscript denotes the initial~(1) and final~(2)
charge states, the first subscript is the number of phonons emitted, and
the second subscript (if any) is the phonon mode $\alpha$ in which they
were emitted.  The frequency of the phonon mode $\alpha$ is given by
$\omega_{\alpha}$ and $\epsilonv_{\alpha}^{(i)}$ is the orthonormal eigenvector of mode $\alpha$ for
the molecule in the electronic state $i$.  
 
The overlap between the two ground vibrational states is
\begin{eqnarray}
\label{GroundStateOverlap}
O_{0,0} &=& \int\,d{\bf{y}} {\Psi^{1}_0}^{*}({\bf{y}}) \Psi^{2}_0({\bf{y}}) \cr
&=& \int\,d{\bf{y}}\Big\{ N_{1} N_{2} 
	\exp\left(-{\bf{y}}^{\dagger} {\mathsf{\Omega}_{1} \over{2\hbar}}
{\bf{y}}\right)\cr 
   & &\times\exp\left(-({\bf{y}}-{\bf{\Delta}})^{\dagger} {\mathsf{\Omega}_{2}\over{2\hbar}}({\bf{y}}-{\bf{\Delta}})\right)\Big\}.
\end{eqnarray}

We now rewrite expression~(\ref{GroundStateOverlap}) so that it contains a
single gaussian rather than a product of two gaussians:
\begin{eqnarray}
\label{singlegaussian}
N_{1}N_{2}\int{d{\bf{y}}}[e^{-{1\over{2\hbar}}{\bf{y}}^{\dagger}
\mathsf{\Omega}_{1}{\bf{y}}}e^{-{1\over{2\hbar}}({\bf{y}}-{\bf{\Delta}})^{\dagger}\mathsf{\Omega}_{2}({\bf{y}}-{\bf{\Delta}})}]\cr
=\int{d{\bf{y}}}[e^{-{1\over{2\hbar}}({\bf{y}}^{\dagger}\mathsf{\Omega}_{1}
{\bf{y}}+{\bf{y}}^{\dagger}\mathsf{\Omega}_{2}{\bf{y}}-{\bf{y}}^{\dagger}
\mathsf{\Omega}_{2}{\bf{\Delta}}-{\bf{\Delta}}^{\dagger}\mathsf{\Omega}_{2}
{\bf{y}}+{\bf{\Delta}}^{\dagger}\mathsf{\Omega}_{2}{\bf{\Delta}})}].
\end{eqnarray} 

\begin{figure}[htb]
\begin{center}
\leavevmode
\unitlength1cm
\begin{picture}(12,7)
\put(-6,-7.4){\makebox(12,12){
\includegraphics{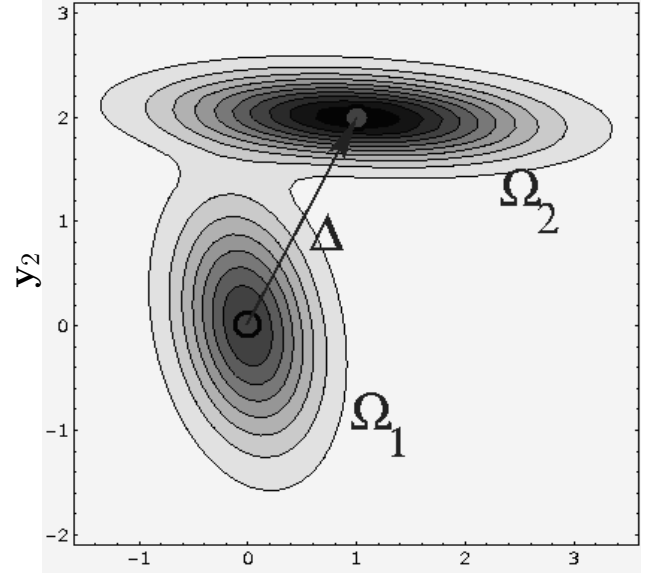}
}}
\setbox6=\hbox{\Large ${\bf{y}}_2$}
\put(0.2,3.7) {\makebox(0,0){\rotl 6}}
\put(4.2,-.7){\Large ${\bf{y}}_1$}
\end{picture}
\end{center}
\caption{Wavefunctions $\Psi^{(1)}_0$ and $\Psi^{(2)}_0$, for harmonic
potentials $\Omega_1^2$ and $\Omega_2^2$ in terms of a two-dimensional rescaled
coordinate $({\bf{y}}_1, {\bf{y}}_2)$, separated by the rescaled length 
$\Delta=\sqrt{m}({\bf{r}}_2-{\bf{r}}_1)$.}
\label{GaussiansA}
\end{figure}

\begin{figure}[htb]
\begin{center}
\leavevmode
\unitlength1cm
\begin{picture}(12,7)
\put(-6,-7.4){\makebox(12,12){
\includegraphics{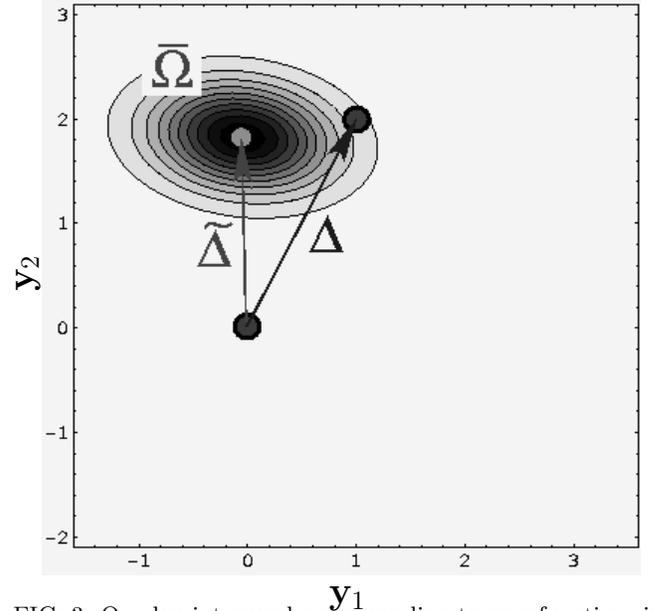}
}}
\setbox6=\hbox{\Large ${\bf{y}}_2$}
\put(0.2,3.7) {\makebox(0,0){\rotl 6}}
\put(4.2,-.7){\Large ${\bf{y}}_1$}
\end{picture}
\end{center}
\caption{Overlap integrand corresponding to wavefunctions in
fig~(\ref{GaussiansA}), centered on $\tilde{\Delta}$ with 
quadratic form $\bar{\Omega}^2$.}
\label{GaussiansB}
\end{figure}

We want to express the single gaussian as one \inThesis{(figure (\ref{GaussiansA}), \ref{GaussiansB})} that is centered on a new origin
${\bf{y}}_{max}={\bf{\tilde{\Delta}}}$ with a new Hessian 
$\mathsf{\bar{\Omega}}$ so that 
the integral is of the form:
  \begin{equation}
  \label{gaussianform}
  N_1 N_2 \int{d({\bf{y}}-\tilde{\bf{{\Delta}}})}
	e^{-{{1\over{\hbar}}
	({\bf{y}}-\tilde{{\bf{\Delta}}})^{\dagger}\mathsf{\bar{\Omega}}
({\bf{y}}-\tilde{{\bf{\Delta}}})}+B}
  \end{equation}
which we know how to solve.  Here $B$ is one of the unknowns we're solving for.
It's a constant which will be pulled out of the integral with a value given in
equation (\ref{Beq}).
\inThesis{
First we find $\vec{y}_{max}$ by taking the derivative of the exponent with
respect to $\vec{y}$ and setting the result equal to 0. 
  \begin{eqnarray}
  {\partial\over{\partial{y}_{i}}}(-2\hbar*exponent)
    &=&[y^{T}(\Omega_{1}+\Omega_{2})]_{i}+[(\Omega_{1}+\Omega_{2})y]_{i}\cr
    & & -(\Omega_{2}\Delta)_{i}-(\Delta^{T}\Omega_{2})_{i}=0.
  \end{eqnarray}
But since $\Omega_{1}$ and $\Omega_{2}$ are symmetric matrices,
$(v^{T}\Omega)_{i}=(\Omega v)_{i}$, this allows us to write: 
  \begin{eqnarray}
  \label{tildadelta}
  2(\Omega_{1}+\Omega_{2})y&=&2\Omega_{2}\Delta\cr
  y_{max}=\tilde{\Delta}&=&(\Omega_{1}+\Omega_{2})^{-1}\Omega_{2}\Delta.
  \end{eqnarray}

Next, we need to find what $\bar{\Omega}$ and $B$ must be by comparing the
exponent in expression~(\ref{gaussianform}) 
with the exponent in expression
(\ref{singlegaussian}) and making sure they are equivalent.
The terms quadratic in $y$ must agree; determining $\bar{\Omega}$ as the
average of the initial and final frequency matrices:
  \begin{eqnarray}
  y^{T}\bar{\Omega}y&=&{1\over{2}}{(y^{T}\Omega_{1}y+y^{T}\Omega_{2}y)}\cr
  \bar{\Omega}&=&{1\over{2}}{(\Omega_{1}+\Omega_{2})}.
  \end{eqnarray}
We also see this comparing the quantities sandwiched by $y^{T}$ and $\Delta$:
  \begin{eqnarray}
  y^{T}\bar{\Omega}(\Omega_{1}+\Omega_{2})^{-1}\Omega_{2}\Delta
  &=&{1\over{2}}{(\y^{T}\Omega_{2}\Delta)}\cr
  \bar{\Omega}&=&{1\over{2}}(\Omega_{1}+\Omega_{2}).
  \end{eqnarray}
Finding $B$ is similar.  We take the constant terms from (\ref{gaussianform})
and (\ref{singlegaussian}) and set them equal to find $B$:
  \begin{eqnarray}
  \label{Beq}
  {\tilde \Delta}^T {\bar \Omega} {\tilde \Delta} - B &=& 
	  {1\over{2}} \Delta^T \Omega_{2} \Delta \nonumber\cr
  B&=& - {1\over{2}} \Delta^T \Omega_{2} \Delta 
	  + {\tilde \Delta}^T {\bar \Omega} {\tilde \Delta} \cr
   &=& -{1\over{2}}{(\Delta^{T}\Omega_{2}\Delta)}
	  +{1\over{2}}\Delta^{T}\Omega_{2}(\Omega_{1}+\Omega_{2})^{-1}
	  \Omega_{2}\Delta\nonumber.
  \end{eqnarray}}

\inPaper{Setting like quantities equal between expressions
(\ref{singlegaussian}) and (\ref{gaussianform}), we obtain 
\begin{eqnarray}
\label{Beq}
\tilde{{\bf{\Delta}}}&=&(\mathsf{\Omega}_1+\mathsf{\Omega}_2)^{-1}
\mathsf{\Omega}_2{\bf{\Delta}}\cr
\mathsf{\bar{\Omega}}&=&{1\over{2}}(\mathsf{\Omega}_1+\mathsf{\Omega}_2)\cr
 B&=& -{1\over{2}}{(\Delta^{T}\Omega_{2}\Delta)}
	  +{1\over{2}}\Delta^{T}\Omega_{2}(\Omega_{1}+\Omega_{2})^{-1}
	  \Omega_{2}\Delta.
\end{eqnarray}
}
Our overlap integral now looks like:
  \begin{eqnarray}
  & &N_1 N_2\int{d{\bf{y}}}
  [e^{-\frac{1}{\hbar}{({\bf{y}}-\tilde{{\bf{\Delta}}})^{\dagger}\mathsf{\bar{\Omega}}({\bf{y}}-\tilde{{\bf{\Delta}}})}}]\cr
& &\times
  e^{-\frac{1}{2\hbar}{{\bf{\Delta}}^{\dagger}\mathsf{\Omega}_{2}
{\bf{\Delta}}}
  e^{\frac{1}{2\hbar}{\bf{\Delta}}^{\dagger}\mathsf{\Omega}_{2}
	(\mathsf{\Omega}_{1}+\mathsf{\Omega}_{2})^{-1}
\mathsf{\Omega}_{2}{\bf{\Delta}}}}.
  \end{eqnarray}
Rewriting the constant part of the integral 
\inThesis{which we found in equation
(\ref{Beq})} in terms of $\tilde{{\bf{\Delta}}}$ and $\mathsf{\bar{\Omega}}$, we have:
  \begin{equation}
  N_1 N_2 \int{d{\bf{y}}}
  [e^{-{{1\over{\hbar}}({\bf{y}}-\tilde{{\bf{\Delta}}})^{\dagger}\mathsf{\bar{\Omega}}({\bf{y}}-\tilde{{\bf{\Delta}}})}}]
  e^{-\frac{1}{2\hbar}{{\bf{\Delta}}^{\dagger}
\mathsf{\Omega}_{2}{\bf{\Delta}}}}
  e^{\frac{1}{\hbar}\tilde{{\bf{\Delta}}}^{\dagger}\bar{\mathsf{\Omega}}
\tilde{{\bf{\Delta}}}}.
  \end{equation}

Changing variables to $\tilde{\bf{y}} ={\bf{ y}} - \tilde {\bf{\Delta}}$, this last integral
is another multidimensional Gaussian, equaling $1/\bar N^2$, where 
$\bar{N}=\sqrt[4]{\det({\mathsf{\bar{\Omega}}\over{\pi\hbar}})}$.
The ground state to ground state overlap is then:
  \begin{equation}
  \label{GroundStateOverlap2}
  O_{0,0} = 
  {N_{1} N_{2} \over \bar{N}^{2}} 
     \exp\left(\frac{1}{\hbar}\tilde{\bf{{\Delta}}}^{\dagger} \mathsf{\bar{\Omega}}\tilde{{\bf{\Delta}}}\right)
	    \exp\left(-{\bf{\Delta}}^{\dagger} \frac{\mathsf{\Omega}_{2}}{2\hbar} {\bf{\Delta}}\right).
  \end{equation}

The probability of being left in the phonon ground state, the
tunneling rate $\Gamma$, and the conductance $G$ are all suppressed by 
a factor $\exp(-g) = |O_{0,0}|^2$, where
\begin{equation}
\label{gtotal1}
G=-\ln(|O_{0,0}|^2).
\end{equation}
This defines the total $g$-factor
which we will use to characterize the overall strength of the phonon coupling.

We can similarly calculate the overlap between the ground initial state
and a final state with one phonon excited into mode $\alpha$:
  \begin{eqnarray}
  \label{FirstExcitedOverlap}
  O_{0,1\alpha} &=& \int\,d{\bf{y}} {\Psi_{0}^{(1)*}({\bf{y}}) 
		\Psi_{1,\alpha}^{(2)}({\bf{y}}-{\bf{\Delta}})} \cr
  &=& \int\,d{\bf{y}}\Big\{ 
      N_{1} e^{-{1\over{2\hbar}} {\bf{y}}^{\dagger}\mathsf{\Omega}_{1}{\bf{y}}}
      N_{2} \, 
        \sqrt{2\omega_\alpha/\hbar} 
        (({\bf{y}}-{\bf{\Delta}})\cdot {\hat{{\epsilonv}}}_{\alpha}^{(2)}) \cr 
   & &\times\exp\left(-{1\over{2\hbar}}
	        ({\bf{y}}-{\bf{\Delta}})^{\dagger} \mathsf{\Omega}_{2} ({\bf{y}}-{\bf{\Delta}})\right)\Big\}.
      \nonumber
  \end{eqnarray}
Combining the exponentials, rewriting them in terms of $\mathsf{\bar \Omega}$ and
$\tilde {\bf{\Delta}}$, we find:
  \begin{eqnarray}
  O_{0,1\alpha} &=&
  N_1 N_2 \int{d{\bf{y}}}\Big\{ 
        \sqrt{2\omega_\alpha/\hbar} 
        (({\bf{y}}-{\bf{\Delta}})\cdot {\hat {{\epsilonv}}}_{\alpha}^{(2)}) \cr 
  & & \times e^{-{{1\over{\hbar}}({\bf{y}}-\tilde{{\bf{\Delta}}})^{\dagger}\mathsf{\bar{\Omega}}({\bf{y}}-\tilde{{\bf{\Delta}}})}}e^{-\frac{1}{2\hbar}{{\bf{\Delta}}^{\dagger}\mathsf{\Omega}_{2}{\bf{\Delta}}}}
  e^{\frac{1}{\hbar}\tilde{{\bf{\Delta}}}^{\dagger}\mathsf{\bar{\Omega}}
\tilde{{\bf{\Delta}}}}\Big\}.
  \end{eqnarray}
Changing variables to $\tilde{\bf{ y}} ={\bf{ y}} - {\tilde{\bf{ \Delta}}}$, we have
  \begin{eqnarray}
  O_{0,1\alpha} &=&
  N_1 N_2 \int{d{\tilde {\bf{y}}}}\Big\{ 
        \sqrt{2\omega_\alpha/\hbar} 
        (({\tilde{\bf{ y}}}-({\bf{\Delta}}-\tilde{\bf{ \Delta}}))\cdot {\hat{\bf{ \epsilon}}}_{\alpha}^{(2)}) \cr 
  & &\times e^{-{{1\over{\hbar}}(\tilde{\bf{ y}})^{\dagger}\mathsf{\bar{\Omega}}(\tilde{\bf{ y}})}}\Big\}
e^{-\frac{1}{2\hbar}{{\bf{\Delta}}^{\dagger}\mathsf{\Omega}_{2}{\bf{\Delta}}}}
  e^{\frac{1}{\hbar}\tilde{\bf{\Delta}}^{\dagger}\mathsf{\bar{\Omega}}\tilde{{\bf{\Delta}}}}\nonumber\cr
  &=&
  N_1 N_2 \int{d{\tilde{\bf{ y}}}}\Big\{ 
        \sqrt{2\omega_\alpha/\hbar} 
        ({\tilde{\bf{ y}}}\cdot {\hat {{\epsilonv}}}_{\alpha}^{(2)}) \, 
  e^{-{{1\over{\hbar}}{\tilde{\bf{ y}}}^{\dagger}\mathsf{\bar{\Omega}}{\tilde{\bf{ y}}}}}\cr
     & &- \sqrt{2\omega_\alpha/\hbar}({\bf{\Delta}} - \tilde{\bf{ \Delta}}) 
	\cdot {\hat {\epsilonv}}_{\alpha}^{(2)}) (1/\bar N)^2
		\Big\} \nonumber\cr
  & &\times e^{-\frac{1}{2\hbar}{{\bf{\Delta}}^{\dagger}\mathsf{\Omega}_{2}{\bf{\Delta}}}}
  e^{\frac{1}{\hbar}\tilde{{\bf{\Delta}}}^{\dagger}\mathsf{\bar{\Omega}}
\tilde{{\bf{\Delta}}}}.
  \end{eqnarray}
Since the first term in the last integral is odd in $\tilde {\bf{y}}$, it must vanish. 

Hence, from
equation~\ref{GroundStateOverlap2}, 
the overlap between the ground initial state and the excited final state is 
  \begin{eqnarray}
  \label{ExcitedStateOverlap}
  O_{0,1\alpha} = O_{0,0} \left(\sqrt{2\omega_{\alpha}/\hbar}
  \left({\hat{{\epsilonv}}}_{\alpha}^{(2)}\cdot 
{(\tilde{{\bf{\Delta}}}-{\bf{\Delta}}})\right)\right).
  \end{eqnarray}

\inThesis{We define:
\begin{eqnarray}
\label{galpha}
g_{\alpha}&=&{{|O_{0,1\alpha}|^2}\over{|O_{0,0}|^2}}\cr
&=&{{P_{\alpha}}\over{P_{ground}}}=
{{\Delta I_{\alpha}}\over{\Delta I_{ground}}},
\end{eqnarray}
which experimentally
gives the ratio of the current flowing emitting one phonon in mode $\alpha$
per electron to the current emitting zero phonons (the ratio of the step
heights in the $dI/dV$ curves: see figure\ref{current}. Thus,
the expression for the g-factor associated with a one-phonon (of mode $\alpha$)
excitation is then given simply by:
\begin{equation}
\label{multidgfactor}
g_{\alpha} = { \left(\sqrt{2\omega_{\alpha}}(q^{(\alpha)} \cdot {\left(\tilde{\Delta}-\Delta\right)})\right)}^{2}.
\end{equation}
}
\inPaper{We define:
\begin{eqnarray}
g_{\alpha}&=&{{|O_{0,1\alpha}|^2}\over{|O_{0,0}|^2}}\cr
&=&{{P_{\alpha}}\over{P_{ground}}}=
{{\Delta I_{\alpha}}\over{\Delta I_{ground}}}
\end{eqnarray}
which experimentally gives the ratio of the current flowing emitting one phonon
in mode $\alpha$ per electron to the current emitting zero phonons (the 
ratio of the step heights in the $dI/dV$ curves).  Thus, 
\begin{equation}
\label{multidgfactor}
g_{\alpha}=\left(\sqrt{2\omega_{\alpha}/\hbar}({{\epsilonv}}_{\alpha}^{(2)}\cdot(\tilde{{\bf{\Delta}}}-{\bf{\Delta}}))\right)^{2}.
\end{equation}}

In\inThesis{the next section, we will discuss} the special case $\mathsf{\Omega}_1=\mathsf{\Omega}_2$,
where the change in charge state does not alter the spring constant
matrices $\mathsf{K}_1$ and $\mathsf{K}_2$, the phonon frequencies and normal
modes remain
unchanged. It is well known that the total overlap
integral is related to the one--phonon emission rates in a simple way:
specifically $G = \sum_\alpha g_\alpha$. This is no longer the case when
the two charge states have different spring constant matrices: we must calculate
them explicitly~\cite{D37}.
The probability of multiple phonons being emitted into {\em distinct}
phonon modes
is given by $g_\alpha g_\beta \dots |O_{0,0}|^2$, as it 
is for the traditionally
studied case $\mathsf{\Omega}_1 = \mathsf{\Omega}_2$. 
But the probability for $n$ phonons
to be emitted into the {\em same} final state is no longer 
$\frac{g_\alpha^n}{n!} |O_{0,0}|^2$. We do the calculation
of two phonons in the Appendix; more general Duschinsky rotation calculations
can be found in the literature~\cite{KRLC93}.

\section{Methods}
\label{sec:Methods}
We used Gaussian 2003, a quantum chemistry package to calculate all of the
quantities needed in our calculation.  These quantities include the
the force constant matrix $K$ for different charge states of the molecule, with dimension $3N 
\times 3N$.  This matrix is related to the $\Omega^2$ matrix by the equation
$K=M\Omega^2$ since in the cases of both $C_{140}$ and $C_{72}$, $M$ commutes with $\Omega$.  
We obtain the vibration frequency eigenvalues and normal mode eigenvectors from
$\Omega^2$.   

The program also gives the 
geometrically minimized structures of the molecule for its different charge
states $r$ and the forces on the
atoms $f$ under the influence of an external electric field.   

All quantities are calculated under the hybrid B3LYP level of theory of the DFT
(density functional theory) model.  The basis set used was STO-3G.  The energy
and minimum geometric structure were also calculated for neutral and charged $C_{72}$ using
the more complete 3-21G* basis set.  Preliminary comparisons with the more
complete basis set suggest that qualitative features are similar to the 
simpler basis set.
All matrix calculations are done under Matlab or its freeware clone GNU Octave.

Because we were working with molecules of considerable size and were calculating
vibrational modes which require many electronic relaxation calculations, we used the minimal
STO-3G basis set for our larger molecule ($N=140$) and the slightly larger 3-21G* basis
set for our smaller molecule ($N=72$).  More complete basis sets would capture the
polarization and charging effects more accurately which would serve to increase
our g factors since the variation between neutral and charged species would be
more pronounced.  However, our analytic approaches and their aim are independent
of the details of the quantum chemistry calculation.
\section{The molecules and their modes}
Our studies were inspired by work done in the McEuen and Ralph groups at Cornell
and Berkeley~\cite{pasupathy1,park1,park2}.  \inThesis{The earliest molecule we looked at was a complex
($[Co(tpy-(CH_{2})_{5}-SH)_{2}]^{2+}$) containing a Cobalt ion
attached to polypyridal ligands and then attached to conjugated polymers
\inPaper{as described in ~\cite{pasupathy1}.}\inThesis{on two
sides as shown in figure~\ref{cobalt}.
\begin{figure}
\vspace{4cm}
\begin{center}
\includegraphics[width=\figWidth]{cobalt.ps}
\end{center}
\caption{$[Co(tpy-(CH_{2})_{5}-SH)_{2}]^{2+}$ molecule.  Electron transport
occurs principally through the orbitals of the Cobalt atom, effectively making
this device a single-atom transistor.}
\vspace{4cm}
\label{cobalt}
\end{figure}
}
We did not use DFT in this initial study.
Electrochemical studies indicate that the Cobalt in this molecule assumes
charges of 2+ and 3+ when a low voltage is applied to it.  Using this
information, the vibrational modes were obtained via Hyperchem 7.0 and the
semi-empirical method ZINDO/S which is an INDO (intermediate neglect of
differential overlap) method parameterized for transition metals.  The modes
found began at around 1 meV and then appeared at intervals of .5 meV which
corresponded approximately to experiments.  We did not calculate 
phonon overlaps
for this molecule, and so we will not discuss it further here. 
The next molecule we worked on was also motivated by an experiment done by the
McEuen and Ralph groups ~\cite{park2}.}Specifically, we looked at the
single molecule transistor made up of $C_{140}$~\cite{pasupathy1}, a molecule whose 
vibrational modes have
been modeled and studied experimentally by Raman 
spectroscopy~\cite{lebedkin}.  $C_{140}$
is comprised of two $C_{70}$ fullerene cages covalently bonded 
to each other via
two $C-C$ bonds\inThesis{ (see figure~(\ref{c140})).  
\begin{figure}
\vspace{4cm}
\begin{center}
\leavevmode
\includegraphics[width=\figWidth]{c140.ps}
\end{center}
\caption{$C_{140}$ coordinates generated by Gaussian DFT}
\label{c140}
\vspace{4cm}
\end{figure}
} 
The dominant mode is the low energy inter-cage vibration stretch mode at 11 meV shown
schematically in figure \inThesis{(\ref{c140mode})}\inPaper{(\ref{c140mod})}.
\inThesis{
\begin{figure}
\vspace{4cm}
\begin{center}
\leavevmode
\includegraphics[width=\figWidth]{mode1c140.ps}
\end{center}
\caption{Schematic depiction of dominant 11 meV stretch mode}
\label{c140mode}
\vspace{4cm}
\end{figure}
}
\inPaper{
\begin{figure}[tbh]
\begin{center}
\leavevmode
\includegraphics[width=\figWidth]{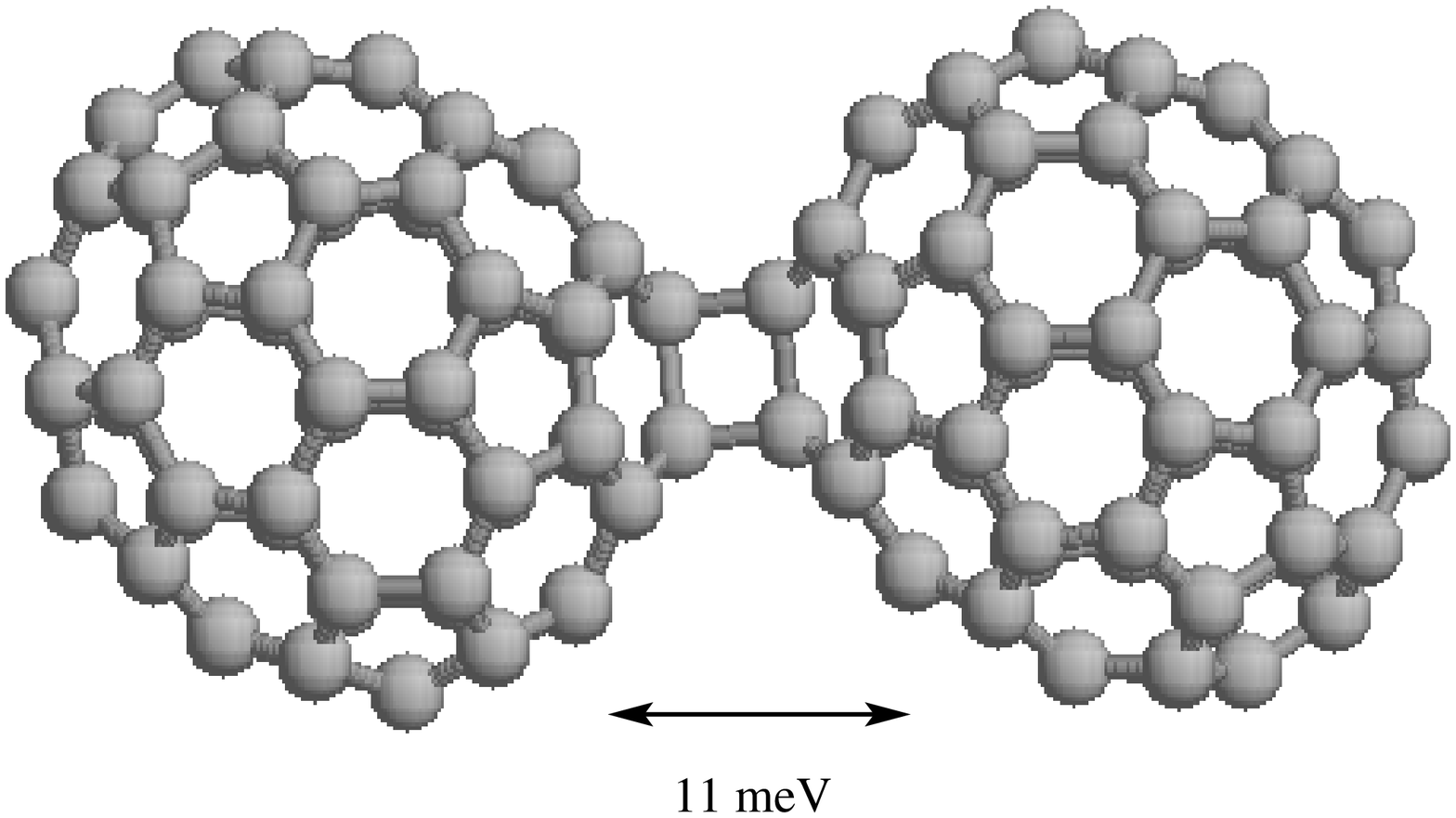}
\end{center}
\caption{$C_{140}$ with stretch mode shown schematically}
\label{c140mod}
\end{figure}
}
The second molecule studied was based on our interest in
$C_{140}$.  We wanted a molecule with similar properties as $C_{140}$, but with
fewer atoms ($C_{72}$).  The aim was to increase the
accuracy of the basis set used for calculations which would be computationally
costly with larger molecules. 

Like $C_{140}$, the dominant mode indicated in our calculations was the
inter-cage stretch mode which has an energy of 19 meV in $C_{72}$.  The molecule is depicted in 
figure~(\ref{c72}).  Figures~(\ref{c140mod}) and~(\ref{c72}) 
were produced using Gaussian2003 to minimize the geometry of the
molecule and RASMOL to plot out the atom positions.

For both molecules, the low energy modes correspond to large scale motion of the
molecules such as the bending, twisting, or stretching of the two cages with
respect to each other (acoustic type vibrations) while higher energy modes
correspond to motion of the atoms on a smaller scale (optical type vibrations).
For example the 15meV mode corresponds to a see-saw motion of the two cages with
respect to each other, the 17meV mode corresponds to a twisting motion of the
two cages away from a central point, while the higher energy 78 meV mode
corresponds to simultaneous deformation of the cages themselves.
The vibrations Gaussian calculate are within $5\%$ of the experimental values. 
\section{Basic Quantities}
\label{sec:Basic}
The shift in the geometrically minimized structure of the $C_{140}$ molecule as it
acquires an extra electron (charge) is the predominant factor in determining the amount of phonon emission.
If the structure changes little, the overlap between the two ground vibrational
states of the initial and final charge state of the molecule will be larger,
which suppresses phonon emission since the overlap is a mathematical statement
of how likely it is for the molecule to remain in the ground vibrational state
rather than transitioning to a excited vibrational state.  

It is not known what the natural charge states of our molecule are on 
a gold substrate, as used in the experiments we compare to. A single
$C_{60}$ molecule typically has charge $-2e$ on gold; doubling this, we
anticipate that the case of interest may involve a transition from perhaps
four to five extra electrons on our molecule.

Table~\ref{table1} is a chart of the change in the inter-cage distance
between the two centers of masses of the
fullerene cages upon adding an electron.
As one can see, the distance increment increases as the charges increases.
\begin{table}
\setlength{\abovecaptionskip}{0pt}
\setlength{\belowcaptionskip}{10pt}
\caption{\label{table1}%
  Change in distances ($\Delta {\bf{r}}$) between centers of mass of the fullerene cages
  for $C_{140}$ during different charge transitions ($Q_1\rightarrow Q_2$) where
  $Q_1$ is the initial charge state of the molecule and $Q_2$ is the final
  charge state of the molecule.
  Shown are the results of our DFT simulations, and those of our simple model 
  (section~\ref{simple}).}
\begin{center}
\begin{tabular}{|c|c|c|}
\hline
Transition $Q_1 \to Q_2$ 
		& DFT $\Delta {\bf{ r}}$ [pm]
			& Simple $\Delta{\bf{r}} = x[Q_2]-x[Q_1]$ \\
\hline
0 $\rightarrow$ 1 & 1.005 & 3.16 \\ \hline
1 $\rightarrow$ 2 &  1.794 & 9.26\\ \hline
2 $\rightarrow$ 3 &  2.333 & 14.8\\ \hline
3 $\rightarrow$ 4 &  3.056 & 19.5\\ \hline
4 $\rightarrow$ 5 &  3.7337 & 23.4 \\ \hline
\end{tabular}
\end{center}
\end{table}
Therefore, as the charge on the molecule increases, the molecular incremental
distortion $\Delta {\bf{r}}$ increases, and consequently the probability that the 
molecule will remain in the
ground vibrational state after an electron has hopped on decreases, leading
to stronger phonon sidebands in the differential conductance graphs.  
Table~\ref{table2}
gives for each electronic transition of the molecule up to a charge
state of 5 extra electrons, the total g-factors (equation~(\ref{gtotal1})) in the
absence of an applied field, the g-factor associated with the first excited
state (eqn~\ref{multidgfactor}) where an inter-cage stretch mode phonon
is emitted, the probability of the molecule remaining in the
ground state ($|O_{0,0}|^2$), and the probability that the molecule's final
vibrational state is the first excited state of the stretch mode
($|O_{0,1\alpha=stretch}|^2$). 
\begin{table}
\vspace{2cm}
\begin{center}
\setlength{\abovecaptionskip}{0pt}
\setlength{\belowcaptionskip}{10pt}
\caption{\label{table2}%
$C_{72}$ undergoing different transitions.  For convenience we include columns 4
and 5; their result can be deduced from the second and third columns}
\begin{tabular}{|c|c|c|c|c|}
\multicolumn{5}{c}{Table: Probabilities and g-factors for different transitions}\\
\hline 
Transition & G & $g_{\alpha=stretch}$&$|O_{0,0}|^2$ & $|O_{0,1\alpha=stretch}|^2$ \\
\hline
0 $\rightarrow$ 1  & 0.960    & $0.33$ &   0.38  &  0.125\\
1 $\rightarrow$ 2  &  1.18 & $0.406$ &  0.31   &    0.126\\
2 $\rightarrow$ 3  & 1.27 & 0.455   &    0.28   &   0.127\\
3 $\rightarrow$ 4  & 1.29  & 0.492   &   0.27   &   0.135\\
\hline
\end{tabular}
\end{center}
\end{table}

\begin{figure}
\begin{center}
\leavevmode
\includegraphics[width=\figWidth]{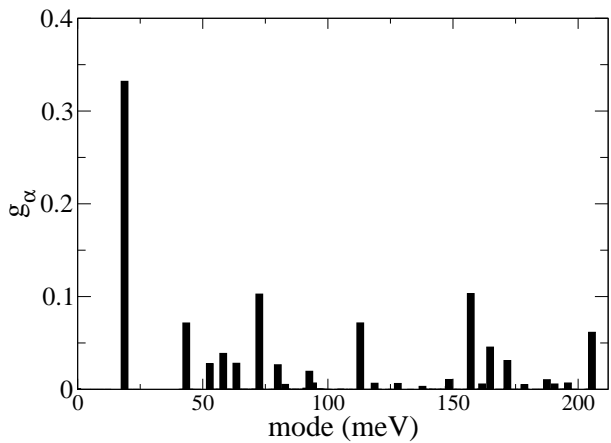}
\end{center}
\caption{$g_{\alpha}$ for the $C_{72}$ $0\to 1$ charge-state transition. The
  large peak is the stretch mode $\alpha=10$ at $19$ meV.  Including the
  two-state emission lines would add an additional peak at $38$ meV (twice the
  stretch mode) and an otherwise roughly continuous background (see Fig.~(\ref{IofVboth}).}
\label{gfactorC72}
\end{figure}

\begin{figure}
\begin{center}
\leavevmode
\includegraphics[width=\figWidth]{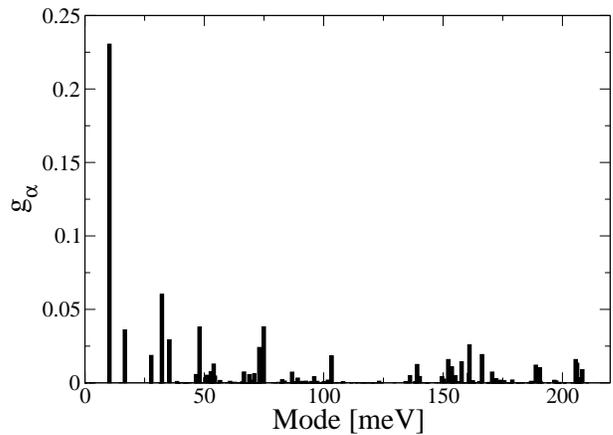}
\end{center}
\caption{$g_{\alpha}$ for the $C_{140}$ $0\to1$ charge-state transition.  The
  large peak is the $\alpha=10$ stretch mode at $11$ meV.  Again,
we estimate that the only significant two-phonon line is at $22$ meV.}
\label{gfactorC140}
\end{figure}

Plotting a graph of the $g_{\alpha}$ factor for the electronic transition of a neutral
molecule to 1- molecule vs. all 216 modes (as in figure~(\ref{gfactorC72})) 
confirms that the stretch mode of the
molecule dominates the single phonon emission. We also plot the 
corresponding graph of $g_{\alpha}$ for $C_{140}$ in
figure~(\ref{gfactorC140}).  As the charge state increases, the effects and
phonon sideband strengths will increase.  Two phonon emission into separate
modes is given by the
product of their respective single mode emission while two phonon emission into
the same mode is given by equation (\ref{GroundStatetoSecondOverlap2}).  In any
case, the approximation we made for equation (\ref{approx40}) suggests that one
phonon emission line given by $g_{11meV}$ dominates.

Again, it is the stretch mode (whose identity is confirmed by displacing 
the equilibrium
coordinates of the molecule by a distortion that is proportional to the
eigenmode) that is important.  Two-phonon
emission may also be significant since experimentally~\cite{pasupathy1} there is sometimes a second smaller
peak at 22meV which may be due to two-phonon emission into the same 11meV mode.
Two-phonon emission, however, yields a small contribution to the conductance.
For two phonons emitted into the same mode, the contribution is given by the
product of the single phonon overlaps.  For two phonons emitted into different
modes, the contribution cannot be simply described by such a product and the
complete expression obtained from integrating the product of the relevant multidimensional
gaussians is needed. 
Although we can calculate the probability of transition to 2-phonon 
up to n-phonon
vibrational final states, we confine ourselves to one-phonon 
emission in our calculations because as will be illustrated in figure
(\ref{IofVboth}) two-phonon emission contributes a continuous background with
the only sizeable contribution from two phonon emission into the 11meV mode.  

\section{Considering External Electric Fields}
\label{sec:Field}  
In reality, the molecule is not in a vacuum but in a real environment of leads
and substrate.  In the experiments~\cite{pasupathy1,park2}, there is a range of
g-factors for different experiments involving the same molecule.  This implies
that environmental effects play an important role and motivates our calculation
of g-factors in the presence of external fields.  We account for one feature of
this variable environment by applying an electric field to
the system.  This external field can come about as a result of image charges
that are set up across the substrate or across the leads when extra electrons
are added to the quantum molecular dot.

\begin{figure}
\begin{center}
\leavevmode
\includegraphics[width=\figWidth]{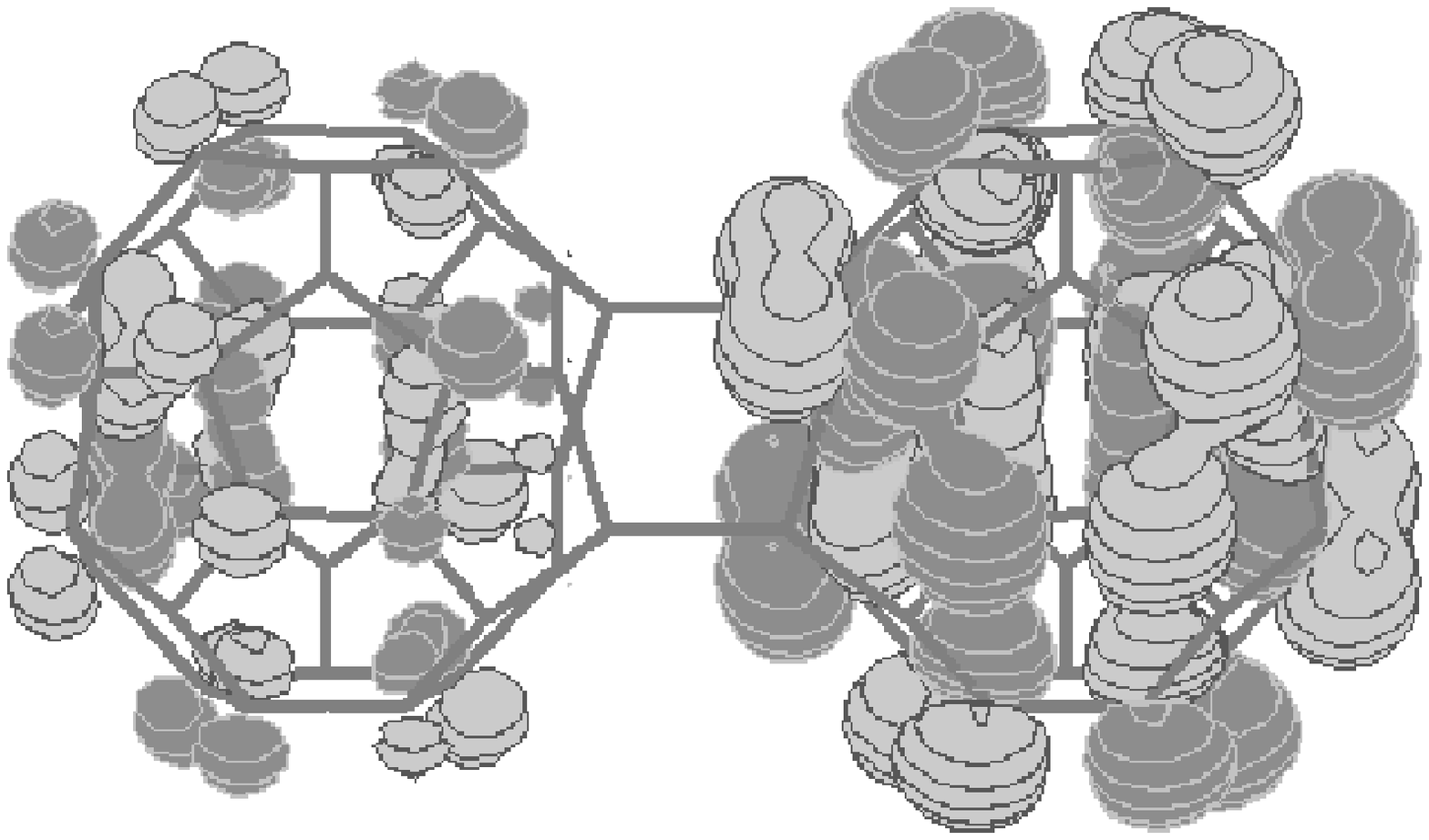}
\end{center}
\caption{$|\Psi|^2$ of the HOMO level of $C_{72}$ under an electric field of  $4\times 10^9$ V/m along
  inter-cage bond, showing the polarization of the electron density}
\label{fig:c72polarize}
\end{figure}

\begin{figure}
\begin{center}
\leavevmode
\includegraphics[width=9cm]{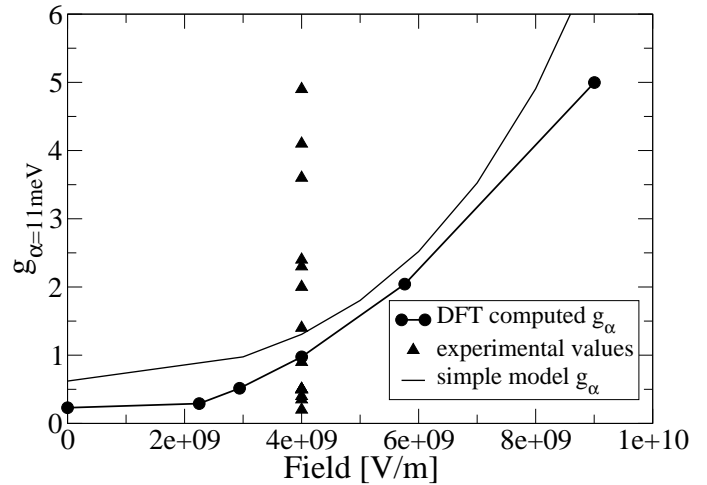}
\end{center}
\caption{Field dependence of $g_{\alpha}$ for the C$_{140}$ $11$ meV stretch
  mode from the DFT calculation.  The solid line is the field dependence for our
  simple model calculation which is explained further in section~\ref{simple}.  Experimental
values (triangles) are taken at zero field, but included on the plot in a vertical column for visibility}
\label{fig:field_dependence2}
\end{figure}

In the Gaussian2003 program, we can impose an external field, relax the
electronic wavefunction due to the induced polarization and measure the 
force (expressed as a $3N$ vector, in this case 216-vector) on each atom.  The
external field will polarize the charge on the molecule as seen in the
following representation in figure~(\ref{fig:c72polarize}) of the highest occupied molecular orbital under the
influence of an external field along the inter-cage bond of the molecule
(rendered using the freeware Molden).
This force will then act to distort the molecule's atomic configuration via
lattice relaxation, leading to an
increase in pathways available to the electron via vibration assisted
tunneling.  The initial and final configurations in equations
(\ref{first}),(\ref{second}),and (\ref{third}) are
${\bf{r}}_1={\bf{r}}_1^{(0)}+\mathsf{K}_{1}^{-1}{\bf{f}}_{1}$ and 
${\bf{r}}_2={\bf{r}}_2^{(0)}+\mathsf{K}_{2}^{-1}{\bf{f}}_{2}$, allowing us to calculate the
$g_{\alpha}$ factors and hence the phonon emission rates from 
equation~(\ref{multidgfactor}).
Figure (\ref{fig:field_dependence2}) shows that $g_{\alpha}$ for the 11meV line
for $C_{140}$
increases substantially under an external field.

As can be seen in the plot, the field does increase the g-factor from its bare
value.  At reasonable fields (those that we might expect to find in 
the experimental
literature) such as the region where the field $\approx 3\times10^9 V/m$, 
$g_{\alpha}$
for the most represented mode (the stretch mode) increases to about 0.5. This
field would correspond to a charge placed 7 angstroms away.  And for a field
corresponding to a charge placed 6 angstroms away (the closest plausible
distance), $g_{\alpha}$ becomes around
1.0.  However, in
experiments, the g-factor varies from values of much less than 1 to values as
high as 6.  In order to reach these quantities in our present theory, we would
need to impose much higher and unphysical fields.  

\inThesis{A subtlety that must be taken into account is taking the inverse of a matrix
which contains zero or near zero elements. The force constant matrix $K$ is such a matrix, but we need
its inverse to calculate how much the equilibrium positions of the atoms
change.  This is problematical because K has zero eigenvalues for the three
translational and two rotational modes.  We use singular value decomposition to tease out the zero frequency
center of 
mass and rotation
modes, fix them, and invert $K$ for the other degrees of freedom.}

\begin{figure}
\leavevmode
\includegraphics[width=9cm]{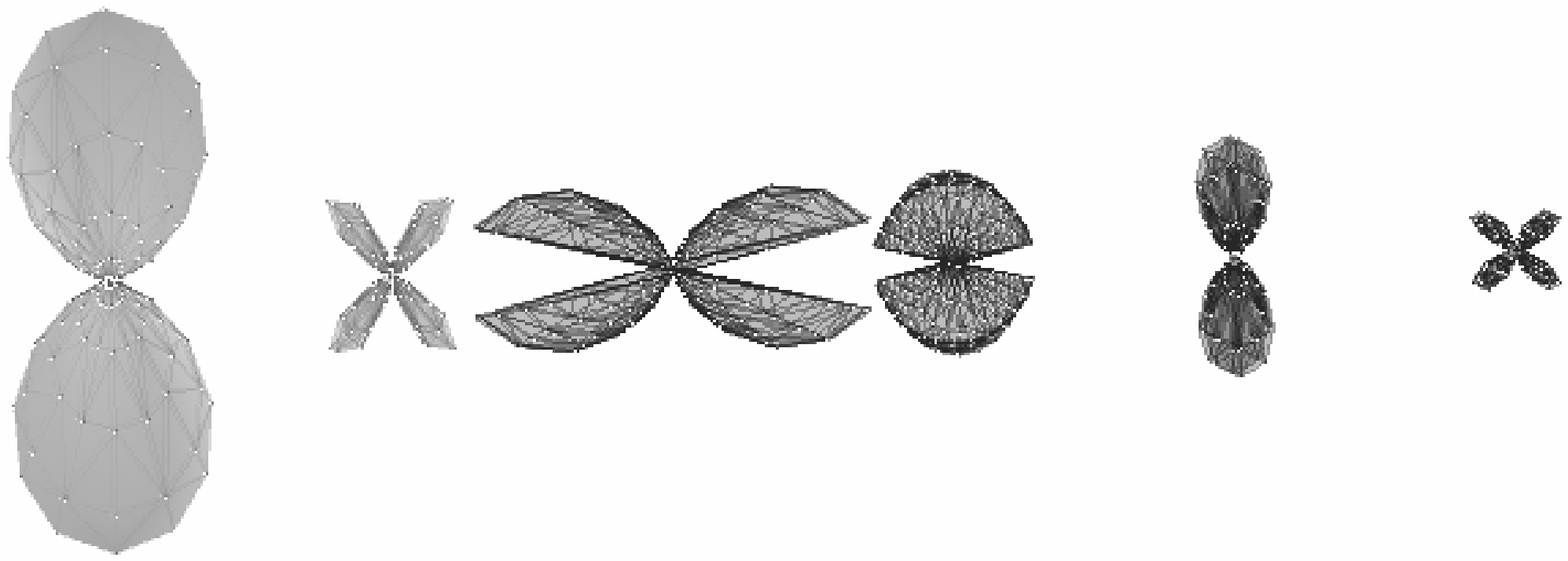}
\caption{Angle Dependence of $g_{\alpha}=stretch$ for $C_{140}$.  Leftmost figure is $g$
plotted as function of the angle of electric field (an extremely high field
magnitude of $5\times 10^{12} V/m$) for the 11meV stretch mode; remaining figures
to the right are for other modes, and have been magnified considerably
to show their shape. The vertical represents fields along the long axis
of the molecule.}
\label{fig:Angle dependence}
\end{figure}

\begin{figure}
\begin{center}
\leavevmode
\includegraphics[width=\figWidth]{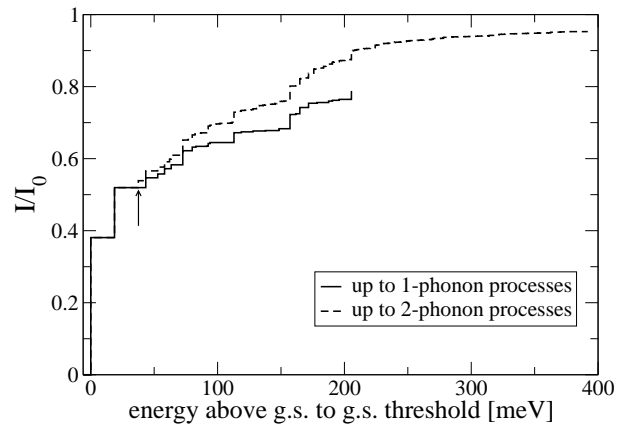}
\end{center}
\caption{I-V curve predicted for $C_{72}$ for one-phonon process (solid line)
and up to two-phonon processes (approximate, dashed line), using
the DFT STO-3G basis set. The arrow indicates the position of the two-phonon contribution from
the stretch mode.}
\label{IofVboth}
\end{figure}

Another dependence we examined was the g-factor dependence of the various modes
on the angle of a fixed electric field.  In figure (\ref{fig:Angle dependence}), the electric field
was fixed at a value of $4\times 10^9$ V/m.  The leftmost figure 
is the 11meV
mode -- the stretch mode.  Following it from left to right are the 3.7 meV mode
magnified by a factor of 20,000; the 2.37 meV mode magnified by a factor of 20;
the 15 meV mode magnified by a factor of 5; the 17 meV mode magnified 
by a factor
of 200 and finally the 27.6 meV mode magnified by a factor of 500.   

The molecule is oriented such that its long axis is aligned vertically. From 
the figure, we see that there is no coupling of the stretch mode (left shape) 
to the
field when the field is aligned in a direction perpendicular to the stretch
mode, and that there is maximum coupling in the direction parallel to 
the direction of the stretch mode.  Note also that $g_{\alpha}$ is
non-zero for $\alpha=$stretch mode even in the absence of a field.  The
symmetries of the plots in figure (\ref{fig:Angle dependence}) reflects the
symmetry of the modes and how they relate to the symmetry of the applied field.
$C_{140}$ has $C_{2h}$ symmetry so it can be generated by a rotation of angle
$\pi$ around a fixed axis and a symmetry on a plane orthogonal to the fixed
axis.  The normal modes of a molecule also possess a definite symmetry with
respect to the planes of symmetry of the molecule. The symmetry of the stretch
mode is even under reflection in the x-y plane, coinciding with the symmetry of
the field $E_{x}$ and orthogonal to the field $E_{z}$.  Thus, the strongest
coupling of the stretch mode is to $E_{x}$.  

\section{A simple model for overlaps and fields}
\label{simple}
To what extent are these quantum overlaps a result of complex quantum chemistry
(bonding and anti-bonding and electronic rearrangements inside the two cages)?
How much can we understand from simple electrostatics of dumbbells?  
\inThesis{Here we pursue further calculations originally made by Ralph and
Pasupathy~\cite{pasupathy1}.}
By modeling the system simply as two rigid balls connected by a spring
subject to an external field, we can obtain some understanding at how the
dimensions of the problem as well as simple quantities might affect the overlap
and g-factor.

We write down the total energy of the system and then minimize the energy with
respect to the parameters of our problem and in the presence of an external
field -- for our case we choose to minimize
the charge on one ball and the distance $x$ between the two balls.

The quantities we take into account are as follows:
\begin{eqnarray}
\label{eqn:Energies}
E_{spring}&=&{1\over{2}} K(x_{2}-x_{1}-a)^2 \cr
E_{field}&=&q_{1} E x_{1}+q_{2}Ex_{2} \cr
E_{coulomb}&=&{Kq_{1}q_{2}\over{(x_{2}-x_{1})}} \cr
E_{capacitance}&=&{{1\over{2}} {kq_{1}^{2}\over{r}}}+{{1\over{2}} {kq_{2}^{2}\over{r}}} 
\end{eqnarray}
where $a$ is the
equilibrium distance of the spring, $x_1$ and $x_2$ are the coordinates of the 
two balls, $r$ is their radius, $K$ is the
spring constant of the system, and $k$ is the Coulomb constant.

\inThesis{First we rewrite our problem for the system in terms $X$, the center 
of mass coordinates; $x_{rel}$; the relative motion coordinates; and $x$, 
the deviation from the
equilibrium distance for a spring:
\begin{eqnarray}
{x_{1}+x_{2}\over{2}}&=&X \cr
x_{2}-x_{1}&=&x_{rel}\cr
x_1&=&X-{{(x+a)}\over{2}}\cr
x_2&=&X+{{(x+a)}\over{2}}
\end{eqnarray}
where $x_{1}$ and $x_{2}$ are the original coordinates of the two balls.
}
We also note that $M_{total}=M_{ball_{1}}+M_{ball_{2}}=2M_{ball}$ and
$M_{red}={{M_{ball_{1}}M_{ball_{2}}}\over{M_{ball_{1}}+M_{ball_{2}}}}=M_{ball}/2$
are the well-known center of mass and reduced mass for the system. 
The last assignment we make are expressions for the charges on each ball ($q_1$
and $q_2$) in
terms of the charges in the system:
\begin{eqnarray}
q_1&=&{Q\over{2}}+{q\over{2}}\cr
q_2&=&{Q\over{2}}-{q\over{2}}
\end{eqnarray}
where $Q$ is the total charge of the system and $q$ is the difference between
the charges on the two balls.
The potential energy U then becomes:
\begin{eqnarray}
U&=&E_{spring}+E_{field}+E_{coulomb}+E_{capacitance}\cr
&=&{1\over{4r(a+x)}}\Big\{-2a^2Eqr+K(q^2(x-r)+Q^2(x+r))\cr
&+&2rx(2EQX-Eqx+kx^2)\cr
&+&a[K(q^2+Q^2)+2r(2EQX-2Eqx+kx^2)]\Big\}.
\end{eqnarray}
Here $x=x_2-x_1$ is the relative distance between the two balls and
$X={x_{1}+x_{2}\over{2}}$ is
the center of mass coordinates of the system.
We next take the derivative of the potential with respect to $q$ the difference
in charges on
the two balls and set the resulting expression (${dU\over{dq}}$) equal to zero.
Solving this expression for $q$ gives us the minimized distribution of charges
on the balls under an external field:
\begin{equation}
\label{qeq}
q={{Er(a+x)^2}\over{K(a-r+x)}}.
\end{equation}  
Similarly, we take the derivative of the potential energy with respect to the
deviation from equilibrium $x$ (${dU\over{dx}}$), set this expression equal to zero, and solve for
$x$.  We keep terms up to second order in $Q$ and $E$ and get:
\begin{equation}
\label{xeq}
x[Q]=AE^2+BQ^2+CE^2Q^2
\end{equation}
where $A$, $B$, and $C$ are given by:
\begin{eqnarray}
\label{subs}
A&=&\left({{2a^2-5ar-3r^2}\over{4kK(a-r)^3}}\right)ra^2\cr
B&=&{{K(a^2+r^2-2ar)}\over{4a^2k(a-r)^2}}\cr
C&=&\left({{2r-a}\over{8k^2(a-r)^3}}\right)r.
\end{eqnarray}

In table~(\ref{table1}), we compare the $\Delta r=x[Q_2]-x[Q_1]$ in the
absence of a field for our the simple model
and the full DFT calculation discussed earlier where $Q_1$ is the total charge
for the initial system and $Q_2$ is the total charge for the final system. The simple
model has between three and six times the distortion of the quantum
chemistry calculation, likely due to a combination of more effective screening
of the Coulomb repulsion between cages and quantum chemistry effects.

The constants from equation~\ref{subs} which define the expression for $x$ in
equation (\ref{xeq}), in combination with the formula for the zero-point motion
$x_{0}=\sqrt{{\hbar\over{M_{red}\omega_{0}}}}$ and the one-dimensional
equivalent of equation (\ref{multidgfactor}) gives us an expression for
$G$:
\begin{equation}
g=G={(x[Q_2]-x[Q_1])^2\over{4x_{0}^2}}
\end{equation}
where in this one-mode limit the log $G$ of the total overlap equals the
one-phonon emission ratio $g$.
Here $M_{red}$ is is equal to ${70
  m_{Carbon}\over{2}}$, $\omega_{0}=\omega_{stretch}=11meV$ and 
the zero-point motion $x_0$ for $C_{140}$ is $2.17$ pm.

Therefore, using these formulas our $g$ factor for the 
$0\to1$ transition of $C_{140}$ is 0.535 (and for $C_{72}$ it is 
0.92).
The complete $3N$-dimensional calculations in the absence of a field
for the same transition yields a $g_\alpha$ for 
the stretch mode of $0.23$ for $C_{140}$ (and 0.33 for $C_{72}$). 
This difference is not as large as one would expect from the difference
in the center--of--mass motions: the 11meV stretch mode incorporates
motions that do not simply change the center of mass separation. In total,
the many-body DFT calculations show a stretch--mode phonon emission about
a factor of three smaller than that predicted from the simple
physical model, and captures the cage size-dependence of $g$,
rather well.

Finally, we compare the field dependence of the $g_{\alpha}$ between the simple
model and the DFT calculation given in figure~(\ref{fig:field_dependence2}).
The field dependence works out quite well.  This simple model could be made more
realistic by incorporating features from the DFT calculation (such as charging
energies), but that would take us beyond our current illustrative goal.


\section{Current due to phonon transitions}
\label{current}
Using the g-factors corresponding to all of the different single phonon modes,
we graphed a current vs. voltage graph for $C_{72}$ using the 
simplified formula
where all the phonons are identical in both charge states of the molecule.
Figure (11) gives the current divided 
by $I_{0}$ vs. the available energy
above the ground state to ground state threshold for both one-phonon emission
processes (solid line) and up to two-phonon processes (dashed line).  The plots
are constructed by iteratively calculating phonon emission from a pool of
available energy.  As energy decreases, less is available for emitting phonons.        
Our $g_{\alpha}$'s make use of the fact that the phonon quadratic forms
$\Omega$ change between different charge states.
As you can see, the currents due to one-phonon processes and for up to two
phonon processes share similar gross features at the beginning such as the jump
in current at the $19$ meV energy mode corresponding to the stretch mode of the
molecule. However they start to deviate as energy increases until 
they level off
at different values of current (0.8 for the one-phonon process and 0.95 for the
two-phonon processes) which would seem to indicate that two-phonon processes
will play a role in the I-V characteristics of a molecular quantum dot.  

In addition, the I-V curve that includes all n-phonon processes will asymptote
to one.  The two-phonon contribution forms almost a continuous background,
except for $2\omega_{stretch}$, whose position is 
shown with an arrow in figure (11).
We also note that the our treatment of two-identical-phonon emission is
(for convenience) not the correct formula derived in equation~(\ref{GroundStatetoSecondOverlap2})
which allows the frequencies to change between the initial and final states, but
the approximate formulas given by equations~\ref{simpleoverlap} and~\ref{iratio}.   
\begin{equation}
\label{simpleoverlap}
O_{0,2\alpha}\approx e^{-G}g_{\alpha}^2/2
\end{equation}
\begin{eqnarray}
\label{iratio}
I/I_0(E)&=&\sum_{\alpha}(0_{0,1\alpha}\Theta(E-\hbar\omega_{alpha})+O_{0,2\alpha}
\Theta(E-2\hbar\omega_{\alpha}))\cr
&+&\sum_{\alpha,\alpha'\neq\alpha}O_{0,1\alpha1\alpha'}\Theta(E-\hbar\omega_{\alpha}-\hbar\omega_{\alpha'})
\end{eqnarray}
As one may observe, the approximate two-phonon rates form a featureless
background except for the 2-stretch-mode phonon peak, a result of the weak
couplings to the other modes.
\section{Conclusion}
\label{conclusion}

There is much recent interest in vibrating mechanical systems coupled to
electron transport on the nanoscale, from nanomechanical 
resonators~\cite{LaHaye,Naik} to single-electron shuttles~\cite{Gorelik, Blick}. 
Vibrational effects on electron transport through molecules have been
studied since the 1960s in devices containing many molecules~\cite{Jaklevic}, 
and more recently have been shown to be important in transport through
single molecules measured using scanning tunneling microscopes~\cite{stipe},
single-molecule transistors~\cite{park1,park2},and mechanical break
junctions~\cite{smit}.
In a natural extension of work done in the 1920s by Franck, Condon, {et~al.}
in atomic spectra, we have studied the effects of molecular vibrations on
electron transport through a molecule. We have shown that density functional
theory calculations
of the normal modes and deformations, coupled to a straightforward linear
algebra calculation, can provide quantitative predictions for the entire
differential tunneling spectrum, even including external fields from
the molecular environment.

\subsection{Acknowledgments}
We would like to thank Jonas Goldsmith and Geoff Hutchinson for helpful
conversations.  We would also like to thank the two referees for their careful
reading of the manuscript and their many helpful suggestions for improvement.  
We acknowledge
support from NSF grants DMR-0218475 and CHE-0403806 and from a GAANN
fellowship DOEd P200A030111.

\section{Appendix}
\label{Appendix}

Here we show how one can calculate the Franck-Condon factor for
a transition from the neutral ground state to an excited state with 
one vibrational mode in a doubly-excited state.
(For emission into general excited states, we would need to use the appropriate
multi-dimensional gaussian multiplied by the appropriate Hermite polynomials.
This calculation quickly becomes complicated~\cite{KRLC93}, and for the
molecules of interest
to us, multiple phonon emission is rare.)
\inThesis{The relevant wavefunctions are:
\begin{eqnarray}
\label{TheStates}
\Psi^{(1)}_{0}({\bf{y}})&=&\NInit e^{-{1\over{2\hbar}} {\bf{y}}^{\dagger} \mathsf{\Omega}_1 {\bf{y}}}\cr
\Psi^{(2)}_{2,\alpha}({\bf{y}}) &=&
N_{2}\sqrt{2}\left(({{\omega_{\alpha}}\over{\hbar}})[\hat{{{\epsilonv}}}_{\alpha}^{(2)}\cdot
({\bf{y}}-{\bf{\Delta}})]^{2}-{1\over{2}}\right)\cr
& & \times e^{-{1\over{2\hbar}}({\bf{y}}-{\bf{\Delta}})^{\dagger}\mathsf{\Omega}_{2}({\bf{y}}-{\bf{\Delta}})}.
\end{eqnarray}
}
\inPaper{From the vibrational states given in (\ref{PsiGroundStates}), we're}
\inThesis{We're} interested in the following integral:
  \begin{equation}
  \label{excitedoverlap}
  O_{0,2} = \int\,d{\bf{y}} {\Psi_{0}^{(1)*}({\bf{y}}) 
 \Psi_{2,\alpha}^{(2)}({\bf{y}}-{\bf{\Delta}})} 
  \end{equation}
where we can split the integral into two parts:
  \begin{align}
  \label{align1}
  & &=\int\,d{\bf{y}} \NInit \NFinal \sqrt{2} 
	\omega_{\alpha}[{{\epsilonv}}_{\alpha}^{(2)}\cdot{({\bf{y}}-{\bf{\Delta}})}]^{2}e^{-{1\over{2\hbar}}\left({\bf{y}}^{\dagger}
	\mathsf{\Omega}_1 {\bf{y}}\right)}\cr
& & \times e^{-{1\over{2\hbar}}\left(({\bf{y}}-{\bf{\Delta}})^{\dagger} \mathsf{\Omega}_2
	({\bf{y}}-{\bf{\Delta}})\right)} \\
  \label{align2}
 & &  -\int\,d{\bf{y}} \NInit \NFinal \sqrt{2}{1\over{2}}e^{-{1\over{2\hbar}}\left({\bf{y}}^{\dagger}
	\mathsf{\Omega}_1 {\bf{y}}\right)} e^{-{1\over{2\hbar}}\left({\bf{y}}-{\bf{\Delta}})^{\dagger} \mathsf{\Omega}_2 ({\bf{y}}-{\bf{\Delta}})\right)}.
  \end{align}

Expression~(\ref{align2}) is just $-{1\over{\sqrt{2}}}O_{0,0}$; we concentrate
on expression~(\ref{align1}).  First, as we did for the $O_{0,1}$ case, we rewrite this integral in terms of the quantities
$\tilde{{\bf{\Delta}}}$ and $\mathsf{\bar{\Omega}}$:
\begin{eqnarray}
\label{excitedo1}
& &N_{1}N_{2}\sqrt{2}({\omega_{\alpha}\over{\hbar}})\int{d{\bf{y}}}[\hat{{{\epsilonv}}}_{\alpha}^{(2)}\cdot
  ({\bf{y}}-{\bf{\Delta}})]^{2}e^{-{1\over{\hbar}}({\bf{y}}-\tilde{{\bf{\Delta}}})^{\dagger}\mathsf{\bar{\Omega}}({\bf{y}}-\tilde{{\bf{\Delta}}})}\cr
& &\times e^{-{1\over{2\hbar}}{{\bf{\Delta}}^{\dagger}\mathsf{\Omega}_{2}{\bf{\Delta}}}}e^{{1\over{2\hbar}}{\tilde{{\bf{\Delta}}}(\mathsf{\Omega}_{1}+\mathsf{\Omega}_{2})\tilde{{\bf{\Delta}}}}}.
\end{eqnarray}

We want to make this expression look like the known gaussian integral:
$C_{1}\int {dx\,x^2 e^{-x^2+C_{2}}}$ where $C_{1}$ and $C_{2}$ are constants.
Changing variables to $\tilde{{\bf{y}}}$:
  \begin{eqnarray}
  \tilde{{\bf{y}}}={\bf{y}}-\tilde{{\bf{\Delta}}}\cr
  d\tilde{{\bf{y}}}=d{\bf{y}}\cr
  {\bf{y}}=\tilde{{\bf{y}}}+\tilde{{\bf{\Delta}}}
  \end{eqnarray}
we rewrite the integral as one over $d^{n}\tilde{y}$:
  \begin{eqnarray}
  \label{excitedo2}
& &N_{1}N_{2}\sqrt{2}({\omega_{\alpha}\over{\hbar}})\int{d\tilde{{\bf{y}}}}[\hat{{{\epsilonv}}}_{\alpha}^{(2)}\cdot
  (\tilde{{\bf{y}}}+\tilde{{\bf{\Delta}}}-{\bf{\Delta}})]^{2}e^{-{1\over{\hbar}}\tilde{{\bf{y}}}^{\dagger}\mathsf{\bar{\Omega}}\tilde{{\bf{y}}}}\cr
& &\times e^{-{1\over{2\hbar}}{{\bf{\Delta}}^{\dagger}\mathsf{\Omega}_{2}{\bf{\Delta}}}}e^{{1\over{2\hbar}}{\tilde{{\bf{\Delta}}}(\mathsf{\Omega}_{1}+\mathsf{\Omega}_{2})\tilde{{\bf{\Delta}}}}}.
  \end{eqnarray}
Expanding out the term in brackets in equation~(\ref{excitedo2}), we get:
  \begin{eqnarray}
  \label{prefactor}
  [\hat{{{\epsilonv}}}_{\alpha}^{(2)}\cdot \tilde{{\bf{y}}}&+&\hat{{{\epsilonv}}}_{\alpha}^{(2)}\cdot
    (\tilde{{\bf{\Delta}}}-{\bf{\Delta}})]^{2}=[\hat{{{\epsilonv}}}_{\alpha}^{(2)}\cdot \tilde{{\bf{y}}}+{\bf{d}}^{2}]^{2}\cr
  &=&(\hat{{{\epsilonv}}}_{\alpha}^{(2)}\cdot \tilde{{\bf{y}}})^{2}+2{\bf{d}}(\hat{{{\epsilonv}}}_{\alpha}^{(2)}\cdot \tilde{{\bf{y}}})+{\bf{d}}^{2}
   \end{eqnarray}
where $d=\hat{{{\epsilonv}}}_{\alpha}^{(2)}\cdot (\tilde{\bf{\Delta}}-{\bf{\Delta}})$.

The second term in equation~(\ref{prefactor}) will be zero in the integral because of
symmetry considerations which dictate that odd powered 
gaussian integrals of the
form: $\int\,dx\, x^{n}e^{-x^{2}}$ where $n$ is odd always
equal zero..  The only terms in the integral of equation~(\ref{excitedo2}) that remain are the first term and the
constant ${\bf{d}}^{2}$.

We transform this integral into the appropriate normal mode basis.  
Since, we are
integrating over the coordinates centered on $\tilde{\bf{\Delta}}$ for a system with a
Hessian of $\mathsf{\bar{\Omega}}$, we want to rewrite everything in terms of the
eigenmodes of the averaged $\mathsf{\bar{\Omega}}$.  We'll call these eigenmodes 
$\hat{{{\rhov}}}_{\beta}$
where the following definitions hold:
  \begin{eqnarray}
  \label{rhodef}
  \tilde{{\bf{y}}}&=&\sum_{\beta}p_{\beta}\hat{{{\rhov}}}_{\beta}\cr
  \mathsf{\bar{\Omega}}\hat{{{\rhov}}}_{\beta}&=&\bar{\omega}_{\beta}\hat{{{\rhov}}}_{\beta}. 
  \end{eqnarray}
Here, $\hat{{{\rhov}}}_{\beta}$ are the orthonormal eigenvectors for $\mathsf{\bar{\Omega}}$
and $p_{\beta}$ are the weightings of each mode's contribution to
$\tilde{{\bf{y}}}$. Hence:
  \begin{eqnarray}
  (\hat{\epsilonv}_{\alpha}^{(2)}\cdot
  \tilde{{\bf{y}}})^{2}&=&(\sum_{\beta}p_{\beta}\hat{\epsilonv}_{\alpha}^{(2)}\cdot \hat{\rhov}_{\beta})^{2}\cr
   &=&\sum_{\beta}p_{\beta}^{2}(\hat{\epsilonv}_{\alpha}^{(2)}\cdot
  \hat{\rhov}_{\beta})^{2}\cr
& &+\sum_{\beta\neq
  \beta^{'}}p_{\beta}p_{\beta^{'}}(\hat{\epsilonv}_{\alpha}^{(2)}\cdot
  \hat{\rhov}_{\beta})(\hat{\epsilonv}_{\alpha}^{(2)}\cdot 
\hat{\rhov}_{\beta^{'}}).
   \end{eqnarray}
Again, the second term is odd in the new integration variables $p_{\beta}$ and
will be zero.
Rewriting the integral in $d\vec{p}$ and remembering that $\hat{\rho}$
diagonalizes $\bar{\Omega}$, the integral from equation~(\ref{excitedoverlap}) 
and~(\ref{excitedo2}) becomes:
  \begin{eqnarray}
  \label{excitedo3}
& &N_{1}N_{2}\sqrt{2}({\omega_{\alpha}\over{\hbar}})[\int{d^{n}p}\sum_{\beta}p_{\beta}^{2}(\hat{\epsilonv}_{\alpha}^{(2)}\cdot\hat{\rhov}_{\beta})^{2}e^{-{1\over{\hbar}}\sum_{\beta^{'}}p_{\beta^{'}}^{2}\omega_{\beta^{'}}}]\cr
& &\times e^{-{1\over{2\hbar}}{{\bf{\Delta}}^{\dagger}\mathsf{\Omega}_{2}{\bf{\Delta}}}}e^{{1\over{2\hbar}}{\tilde{{\bf{\Delta}}}(\mathsf{\Omega}_{1}+\mathsf{\Omega}_{2})\tilde{{\bf{\Delta}}}}}\cr
&=&N_{1}N_{2}\sqrt{2}({{\omega_{\alpha}}\over{\hbar}})\sum_{\beta}(\hat{\epsilonv}_{\alpha}^{(2)}\cdot\hat{\rhov}_{\beta})^{2}[\int{d^{n}p}p_{\beta}^{2}e^{-{1\over{\hbar}}\sum_{\beta^{'}}p_{\beta^{'}}^{2}\bar{\omega}_{\beta^{'}}}]\cr
& &\times e^{-{1\over{2\hbar}}{\bf{\Delta}}^{\dagger}\mathsf{\Omega}_{2}
{\bf{\Delta}}}e^{{1\over{2\hbar}}{\tilde{{\bf{\Delta}}}(\mathsf{\Omega}_{1}
+\mathsf{\Omega}_{2})\tilde{{\bf{\Delta}}}}}.
\end{eqnarray}
But $\int x^{2}e^{-Ax^{2}}dx={\sqrt{\pi}\over{2A^{3/2}}}={1\over{2A}}\int
e^{-Ax^{2}}dx$, so
\begin{eqnarray}
\int{d^{n}p}\,
p_{\beta}^{2}\,e^{-{1\over{\hbar}}\sum_{\beta^{'}}p_{\beta^{'}}^{2}\bar{\omega}_{\beta^{'}}}
&=&{1\over{2\bar{\omega}_{\beta}/\hbar}}\sqrt{{{\pi\hbar}\over{\bar{\omega}_{\beta}}}}\prod_{\beta^{'}\neq\beta}
\sqrt{{\pi\hbar}\over{\bar{\omega}_{\beta^{'}}}}\cr
&=& \frac{\hbar}{2\bar\omega_\beta} \frac{1}{\bar N^2}.
\end{eqnarray}
Hence, the first term in (\ref{prefactor}) from equation~(\ref{excitedo3})
becomes:
\begin{equation}
N_{1}N_{2}\sqrt{2}({{\omega_{\alpha}}\over{\hbar}})\sum_{\beta}(\hat{\epsilonv}_{\alpha}^{(2)}\cdot
  \hat{\rhov_{\beta}})^{2}{{\hbar}\over{2\bar{\omega}_{\beta}}}{1\over{\bar{N}^{2}}}e^{{1\over{\hbar}}\tilde{{\bf{\Delta}}}^{\dagger}\mathsf{\bar{\Omega}}
\tilde{{\bf{\Delta}}}}e^{-{1\over{2\hbar}}{\bf{\Delta}}^{\dagger}
\mathsf{\Omega}_{2}{\bf{\Delta}}}
\end{equation}
which from equation~(\ref{GroundStateOverlap2}) we see is 
\begin{equation}
\label{GroundStatetoSecondOverlap1}
O_{0,0}\sqrt{2}\omega_{\alpha}\sum_{\beta}{1\over{2\bar{\omega}_{\beta}}}(\hat{\epsilonv}_{\alpha}^{(2)}\cdot\hat{\rhov}_{\beta})^{2}.
\end{equation}
Combining this with the third term from equation~(\ref{prefactor}) 
and expression~(\ref{align2}), our expression for the 
$0\rightarrow 2$ overlap becomes:
\begin{eqnarray}
\label{GroundStatetoSecondOverlap2}
O_{0,2_{\alpha}}&=&O_{0,0}\Big\{\sqrt{2}\omega_{\alpha}\sum_{\beta}{1\over{2\bar{\omega}_{\beta}}}(\hat{\epsilonv}_{\alpha}^{(2)}\cdot\hat{\rhov}_{\beta})^{2}
\cr
&+&\sqrt{2}({\omega_{\alpha}\over{\hbar}})[\hat{\epsilonv}_{\alpha}^{(2)}\cdot(\tilde{{\bf{\Delta}}}-{\bf{\Delta}})]^{2}-{1\over{\sqrt{2}}}\Big\}\cr
&=&O_{0,0}\Big\{\sum_{\beta}{\omega_{\alpha}\over{\sqrt{2}\bar{\omega}_{\beta}}}(\hat{\epsilonv}_{\alpha}^{(2)}\cdot\hat{\rhov_{\beta}})^{2}\cr
&+&\sqrt{2}({\omega_{\alpha}\over{\hbar}})[\hat{\epsilonv}_{\alpha}^{(2)}\cdot(\tilde{{\bf{\Delta}}}-{\bf{\Delta}})]^{2}-{1\over{\sqrt{2}}}\Big\}.
\end{eqnarray}

If $\mathsf{\Omega_1}=\mathsf{\Omega}_2$ (i.e., there is no change in the harmonic potential),
$\mathsf{\bar{\Omega}}=\mathsf{\Omega}_{2}$ and hence
$\hat{\epsilonv}_{\alpha}^{(2)}\cdot\hat{\rhov}_{\beta}=\delta_{\alpha\beta}$ 
and
$\bar{\omega}_{\beta}=\omega_{\beta}$.  The first sum reduces to
${1\over{\sqrt{2}}}$, canceling the last term.  Therefore in this case:
\begin{equation}
\label{approx40}
|{O_{0,2\alpha}\over{O_{0,0}}}|^{2}=2{({\omega_{\alpha}\over{\hbar}})}^{2}[\hat{\epsilonv}_{\alpha}^{(2)}\cdot(\tilde{{\bf{\Delta}}}-{\bf{\Delta}})]^{4}={g_{\alpha}^{2}\over{2}}\inPaper{.}
\end{equation}
\inThesis{as we will see again in equation~(\ref{garray}).}

The change in harmonic potential upon charging the molecule allows for phonon
emission even in the absence of a configurational shift.  Therefore, even if
${\bf{\Delta}}=\tilde{\bf{\Delta}}=0$, phonons can be emitted both because of 
frequency shifts
$(\omega_{\alpha}\neq\bar{\omega}_{\beta})$ and because the normal modes change $(\hat{\epsilonv}_{\alpha}\neq\hat{\rhov}_{\beta})$. 

The other overlaps could be performed in a similar way.  The strategy 
is to write
everything in terms of integrals of $\int dx x^{n}e^{-x^{2}}$ by 
transforming to
the basis of the averaged gaussian with $\bar{\Omega}$.  The odd powered
integrals are eliminated and the even powered terms remain.  
\inThesis{
\subsection{1D case}
\label{1Dcase}
In one dimension, the vibrational wavefunctions are:
\begin{eqnarray}
\label{eqn:OneDimPsis}
\psi_0(x) &=& \left({m\omega\over\pi\hbar}\right)^{1/4} 
		e^{-m\omega x^2/2\hbar}\cr
\psi_1(x) &=& \left({m\omega\over\pi\hbar}\right)^{1/4} 
		\sqrt{2m\omega/\hbar} \, x \, e^{-m\omega x^2/2\hbar}\cr
\psi_n(x) &=& \left({m\omega\over\pi\hbar}\right)^{1/4} 
	\left(H_n(\sqrt{m\omega/\hbar} \, x)\over 2^{n/2} \sqrt{n!}\right) 
		e^{-m\omega x^2/2\hbar}.
\end{eqnarray}
The overlaps in one-dimension are then:
\begin{eqnarray}
\label{eqn:OneDimOverlaps00}
O_{0,0} &=& \int d x \left(\frac{m \omega_{1}}{\pi\hbar}\right)^{1/4}
              e^{-m\omega_{1} x^2/2\hbar}
                         \left({m\omega_{2}\over\pi\hbar}\right)^{1/4}
              e^{-m\omega_{2}(x-a)^2/2\hbar} \cr
        &=& \frac{\left(\frac{m \omega_{1}}{\hbar}\right)^{1/4}
            \left(\frac{m \omega_{2}}{\hbar}\right)^{1/4}}
	      {\left(\left(\frac{m
              (\omega_{1}+\omega_{2})}{2\hbar}\right)^{1/4}\right)^2} 
             e^{-{m\over{2 \hbar}}{\omega_{1}\omega_{2}a^{2}\over{(\omega_{1}+\omega_{2})}}}
\end{eqnarray}
\begin{eqnarray}
\label{eqn:OneDimOverlaps01}
O_{0,1} &=& \frac{\left(\frac{m \omega_{1}}{\hbar}\right)^{1/4}
            \left(\frac{m \omega_{2}}{\hbar}\right)^{1/4}}
	      {\left(\left(\frac{m
              (\omega_{1}+\omega_{2})}{2\hbar}\right)^{1/4}\right)^2} 
             e^{-{m\over{2
            \hbar}}{\omega_{1}\omega_{2}a^{2}\over{(\omega_{1}+\omega_{2})}}}
            (-\sqrt{{2m\omega_{2}\over{\hbar}}} a) \cr
        &=& O_{0,0} (-\sqrt{{2m\omega_{2}\over{\hbar}}} a)
\end{eqnarray}
\begin{eqnarray}
\label{eqn:OneDimOverlaps02}
O_{0,2} &=& \frac{\left(\frac{m \omega_{1}}{\hbar}\right)^{1/4}
            \left(\frac{m \omega_{2}}{\hbar}\right)^{1/4}}
	      {\left(\left(\frac{m
              (\omega_{1}+\omega_{2})}{2\hbar}\right)^{1/4}\right)^2} 
             e^{-{m\over{2
            \hbar}}{\omega_{1}\omega_{2}a^{2}\over{(\omega_{1}+\omega_{2})}}}
            {{[\sqrt{2}(2a^{2}m\omega_{1}^{2}\omega_{2}+(\omega_{2}^{2}-\omega_{1}^{2})\hbar]}\over{{\hbar(\omega_{1}^{2}+\omega_{2}^{2})}}} \cr
        &=& O_{0,0} {{[\sqrt{2}(2a^{2}m\omega_{1}^{2}\omega_{2}+(\omega_{2}^{2}-\omega_{1}^{2})\hbar]}\over{{\hbar(\omega_{1}^{2}+\omega_{2}^{2})}}}
\end{eqnarray}
where a is the shift in the equilibrium position of the molecule with the
addition of the extra charge or in other words, the displacement of 
the harmonic
oscillator minima.  The g factor which will be discussed in the next 
section, is a measurement of the ratio of the conductance
when one phonon is present versus the conductance when no phonons are 
present is
then just 
\begin{eqnarray}
\label{gfactor}
g=({2m\omega_{2}\over{\hbar}})a^{2}.
\end{eqnarray}

The 3N-dimensional case reduces down to the one-dimensional case which can be
seen by comparing equation~(\ref{gfactor}) to the expression calculated for the
3N-dimensional case equation~(\ref{multidgfactor}) where
$(q^{(\alpha)}\cdot(\tilde{\Delta}-\Delta))\rightarrow {\sqrt{m}}a$
and $\omega_{\alpha}$ is the frequency of the mode $\alpha$ belonging to the
final molecule's excited vibrational state, corresponding to $\omega_{2}$, the frequency of the
ending molecule's vibrational excited state. 
}
\inThesis{
\subsection{g factors and connection with experiments}
\label{subsec:gfactors}
To connect our calculations with experiment, we define a g factor which is a
measure of how much the current is suppressed by the phonon overlap in equation~(\ref{galpha}).

The g factor then gives the current step height in the I vs V. graphs, where $O_{0,\alpha}$ is the phonon overlap governing the transition
from an initial vibrational ground state to an excited vibrational ground state
labeled by $\alpha$. 

A measure of the total g factor is defined by:
\begin{equation}
\label{totalg}
e^{-g}=|O_{0,0}|^{2}
\end{equation}
where $O_{0,0}$ is the overlap for the case when the molecule remains in the
ground vibrational state with the addition of the extra electron.

Considering the 1D problem for the case when $\omega_{1}=\omega_{2}$, when we
expand out $e^{-g}$, we get:
\begin{equation}
\label{gexpansion}
e^{-g}=1-({{m\omega}\over{2\hbar}})a^2+({{m\omega}\over{2\hbar}})^{2}{a^4\over{2!}}-({{m\omega}\over{2\hbar}})^{3}{a^6\over{3!}}+...
\end{equation}
where the terms of the expansion are $\sim g_{\alpha}$ for 1-phonon, 2-phonon,
3-phonon, etc. emission:
\begin{eqnarray}
\label{garray}
g_{\alpha}={|O_{0,1}|^{2}\over{|O_{0,0}|^{2}}}&=&{{m\omega}\over{2\hbar}}a^2\cr
{|O_{0,2}|^{2}\over{|O_{0,0}|^{2}}}&=&({{m\omega}\over{2\hbar}}a^2)^{2}/2!={{g_{\alpha}}^{2}\over{2!}}\cr
{|O_{0,3}|^{2}\over{|O_{0,0}|^{2}}}&=&({{\omega}\over{2\hbar}}a^2)^{3}/3!={{g_{\alpha}}^{3}\over{3!}}.
\end{eqnarray}
\inThesis{
The following is then true:
\begin{eqnarray}
|O_{0,0}|^{2}+|O_{0,1}|^{2}+|O_{0,2}|^{2}+...&=&|O_{0,0}|^{2}(1+{{|O_{0,1}|^{2}}\over{|O_{0,0}|^{2}}}+{{|O_{0,2}|^{2}}\over{|O_{0,0}|^{2}}}+...)\cr
&=&e^{-g}(1+g_{\alpha}+{{g_{\alpha}^{2}}\over{2!}}+{{g_{\alpha}^{3}}\over{3!}}+...)\cr
&=&e^{-g}(e^{+g})=1
\end{eqnarray}
}
and
\begin{equation}
\label{gexpression}
g=\sum_{\alpha}g_{\alpha}.
\end{equation}
It is thus correct to associate the $n^{th}$ term in
expansion~(\ref{gexpansion}) with n-phonon emissions.

For 3N-dimensions, that means that for the case $\Omega_{1}=\Omega_{2}$, the probability that
$n_{\alpha}$ phonons will be emitted is given by:
\begin{equation}
\label{simpleP}
P\{n\alpha\}=e^{-g}\,\Pi_{\alpha}{{g_{\alpha}^{n_{\alpha}}}\over{n_{\alpha}!}}.
\end{equation}

However, for our case where $\Omega_{1}\neq\Omega_{2}$, this is not true in
general.\inThesis{ To see this, we go back to the 1D case and consider 
the situation
where $\omega_{1}\neq\omega_{2}$.  Even for the simplified case where we assume
that \emph{a}, the change in molecular configuration due to the addition 
of a charge
(our 3N-D analog would be $\Delta$), is equal to zero, we still obtain 
2-phonon,
4-phonon,...,2n-phonon emission.

Looking to section (\ref{1Dcase}) at our expression for 1D overlaps for the $0\rightarrow 2$
transition, equation~(\ref{eqn:OneDimOverlaps02}), we see that the $|O_{0,2}|$
overlap for different $\omega$'s, but $a=0$ is:
\begin{equation}
|O_{0,2}|=|O_{0,0}|\sqrt{2}\left({{\omega_{2}^{2}-\omega_{1}^{2}}\over{\omega_{1}^{2}+\omega_{2}^{2}}}\right)
\end{equation}
which is clearly non-zero.  Thus, our simple expression for $g$ given in
equation~(\ref{gexpression}) doesn't hold since $g_{\alpha}$, the g 
for one-phonon
emission is zero when $a=0$, but $g\neq 0$ since there are 
non-zero even-phonon
emissions.
}
Therefore, we must calculate the g factors due to different n-phonon emissions
explicitly with equation~(\ref{galpha}) and the appropriate overlaps
given by equations~(\ref{GroundStateOverlap2}), (\ref{ExcitedStateOverlap}), and (\ref{GroundStatetoSecondOverlap2}).
}


\end{document}